\documentclass[11pt]{article}

\usepackage{microtype}
\usepackage{url,epsfig}
\usepackage{hyperref}
\hypersetup{
    colorlinks=true,
    linkcolor=black,
    citecolor=black,
    filecolor=black,
    urlcolor=black,
}
\usepackage{relsize}
\usepackage{fancyvrb}
\usepackage{amsmath}
\usepackage{amssymb}
\usepackage{graphicx}
\graphicspath{{./fig/}}
\usepackage{geometry} 
\usepackage{sectsty} 
\usepackage{subcaption} 
\usepackage[export]{adjustbox} 
\usepackage[normalem]{ulem} 
\sectionfont{\sffamily\bfseries\upshape\large}
\subsectionfont{\sffamily\bfseries\upshape\normalsize} 
\subsubsectionfont{\sffamily\mdseries\upshape\normalsize}
\makeatletter
\renewcommand\@seccntformat[1]{\csname the#1\endcsname.\quad}
\makeatother\renewcommand{\bibitem}{\vskip 2pt\par\hangindent\parindent\hskip-\parindent}

\makeatletter
\def\@maketitle{%
  \begin{center}%
  \let \footnote \thanks
    {\large \@title \par}%
    {\normalsize
      \begin{tabular}[t]{c}%
        \@author
      \end{tabular}\par}%
    {\small \@date}%
  \end{center}%
}
\makeatother

\title{\bf Practical Bayesian model evaluation using leave-one-out cross-validation and WAIC\footnote{We thank Bob Carpenter, Avraham Adler, Joona Karjalainen, Sean Raleigh, Sumio Watanabe, and Ben Lambert for helpful comments, Juho Piironen for R help, Tuomas Sivula for Python port, and the U.S. National Science Foundation, Institute of Education Sciences, and Office of Naval Research for partial support of this research.}\vspace{.1in}}
\author{Aki Vehtari\footnote{Helsinki Institute for Information Technology HIIT, Department of Computer Science, Aalto University, Finland.} \and Andrew Gelman\footnote{Department of Statistics, Columbia University, New York.} \and Jonah Gabry$^{\ddagger}$}
\date{1 September 2016\vspace{-.2in}}
\begin{document}
\maketitle
\thispagestyle{empty}

\begin{abstract}
Leave-one-out cross-validation (LOO) and the widely applicable information
criterion (WAIC) are  methods for estimating pointwise out-of-sample prediction
accuracy from a fitted Bayesian model  using the log-likelihood evaluated at the
posterior simulations of the parameter values.
LOO and WAIC have various advantages over simpler estimates of predictive error
such as AIC and DIC but are less used in practice because they involve
additional computational steps.  Here we lay out fast and stable computations
for LOO and WAIC that can be performed using existing simulation draws.  We
introduce an efficient computation of LOO using Pareto-smoothed importance
sampling (PSIS), a new procedure for regularizing importance weights. Although
WAIC is asymptotically equal to LOO, we demonstrate that PSIS-LOO is more robust
in the finite case with weak priors or influential observations. As a byproduct
of our calculations, we also obtain approximate standard errors for estimated
predictive errors and for comparison of predictive errors between two models.
We implement the computations in an R package called {\tt loo} and demonstrate
using models fit with the Bayesian inference package Stan.

Keywords:  Bayesian computation, leave-one-out cross-validation (LOO), $K$-fold
cross-valida\-tion, widely applicable information criterion (WAIC), Stan, Pareto
smoothed importance sampling (PSIS)
\end{abstract}

\section{Introduction}
After fitting a Bayesian model we often want to measure its predictive accuracy,
for its own sake or for purposes of model comparison, selection, or averaging
(Geisser and Eddy, 1979, Hoeting et al., 1999, Vehtari and Lampinen, 2002, Ando
and Tsay, 2010, Vehtari and Ojanen, 2012).  Cross-validation and information
criteria are two approaches to estimating out-of-sample predictive accuracy
using within-sample fits (Akaike, 1973, Stone, 1977).  In this article we
consider computations using the log-likelihood evaluated at the usual posterior
simulations of the parameters. Computation time for the predictive accuracy
measures should be  negligible compared to the cost of fitting the model and
obtaining posterior draws in the first place.

Exact cross-validation requires re-fitting the model with different training
sets. Approximate leave-one-out cross-validation (LOO) can be computed easily
using importance sampling (IS; Gelfand, Dey, and Chang, 1992, Gelfand, 1996) but
the resulting estimate is noisy, as the variance of the importance weights can
be large or even infinite (Peruggia, 1997, Epifani et al., 2008). 
Here we propose to use \emph{Pareto smoothed importance sampling}
(PSIS), a new approach that provides a more accurate and reliable estimate by fitting a Pareto
distribution to the upper tail of the distribution of the importance weights.
PSIS allows us to compute LOO using importance weights that would otherwise be
unstable. 

WAIC (the widely applicable or Watanabe-Akaike information criterion; Watanabe,
2010)  can be viewed as an improvement on the deviance information criterion
(DIC) for Bayesian models.  DIC has gained popularity in recent years, in part
through its implementation in the graphical modeling package BUGS
(Spiegelhalter, Best, et al., 2002; Spiegelhalter, Thomas, et al., 1994, 2003),
but it is known to have some problems, which arise in part from not being fully
Bayesian in that it is based on a point estimate (van der Linde, 2005, Plummer,
2008). For example, DIC can produce negative estimates of the effective number
of parameters in a model and it is not defined for singular models. WAIC is
fully Bayesian  in that it uses the entire posterior distribution, and it is
asymptotically equal to Bayesian cross-validation. Unlike DIC, WAIC is invariant
to parametrization and also works for singular models.

Although WAIC is asymptotically equal to LOO, we demonstrate that PSIS-LOO is
more robust in the finite case with weak priors or influential observations. We
provide diagnostics for both PSIS-LOO and WAIC which tell when these
approximations are likely to have large errors and computationally more
intensive methods such as $K$-fold cross-validation should be used.
Fast and stable computation and diagnostics for PSIS-LOO allows safe use of this
new method in routine statistical practice.
As a byproduct of our calculations, we also obtain approximate standard errors
for estimated predictive errors and for the comparison of predictive errors
between two models.

We implement the computations in a package for R (R Core Team, 2016) called
{\tt loo} (Vehtari, Gelman, and Gabry, 2016) and demonstrate using models fit
with the Bayesian inference package Stan (Stan Development Team, 2016a, b).%
\footnote{
The {\tt loo} R package is available from CRAN and
\url{https://github.com/stan-dev/loo}. The corresponding code for
Matlab, Octave, and Python is available at
\url{https://github.com/avehtari/PSIS}.}
All the computations are fast compared to the typical time required to fit the
model in the first place. Although the examples provided in this paper all use
Stan, the {\tt loo} package is independent of Stan and can be used with models
estimated by other software packages or custom user-written algorithms.

\section{Estimating out-of-sample pointwise predictive accuracy using posterior simulations}

Consider data $y_1,\dots,y_n$, modeled as independent given parameters $\theta$;
thus $p(y|\theta)=\prod_{i=1}^n p(y_i | \theta)$.
This formulation also encompasses latent variable models with $p(y_i | f_i, \theta)$, 
where  $f_i$ are latent variables.
Also suppose we have a prior distribution $p(\theta)$, 
thus yielding a posterior distribution $p(\theta|y)$ and a posterior predictive distribution 
$p(\tilde{y}|y)=\int \!p(\tilde{y}_i|\theta)p(\theta|y)d\theta$. To maintain comparability with the given 
dataset and to get easier interpretation of the differences in scale of effective number of parameters, 
we define a measure of predictive accuracy for the $n$ data points taken one at a time:
\begin{eqnarray}
\nonumber \mbox{elpd} &=& \mbox{expected log pointwise predictive density for a new dataset}\\
\label{elpd_2} &=& \sum_{i=1}^n \int p_t(\tilde{y}_i) \log p(\tilde{y}_i|y) d\tilde{y}_i,
\end{eqnarray}
where $p_t(\tilde{y}_i)$ is the distribution representing the true
data-generating process for $\tilde{y}_i$. The $p_t(\tilde{y}_i)$'s are unknown,
and we will use cross-validation or WAIC to approximate (\ref{elpd_2}). In a
regression, these distributions are also implicitly conditioned on any
predictors in the model.
See Vehtari and Ojanen (2012) for other approaches to approximating
$p_t(\tilde{y}_i)$ and discussion of alternative prediction tasks.

Instead of the log predictive density $\log p(\tilde{y}_i|y)$, other utility (or
cost) functions $u(p(\tilde{y}|y),\tilde{y})$ could be used, such as
classification error. Here we take the log score as the default for
evaluating the predictive density (Geisser and Eddy, 1979, Bernardo and Smith,
1994, Gneiting and Raftery, 2007).

A helpful quantity in the analysis is
\begin{eqnarray}\label{lpd}
\nonumber \mbox{lpd} &=& \mbox{log pointwise predictive density}\\
&=& \sum_{i=1}^n \log p(y_i|y)=\sum_{i=1}^n \log\!\int \!p(y_i|\theta)p(\theta|y)d\theta.
\end{eqnarray}
The lpd of observed data $y$ is an overestimate of the elpd for future data
(\ref{elpd_2}). To compute the lpd in practice, we can evaluate the expectation
using draws from $p_{\rm post}(\theta)$, the usual posterior simulations, which
we label $\theta^s$, $s=1,\dots,S$:
\begin{eqnarray}
\nonumber \widehat{\mbox{lpd}} &=&\mbox{computed log pointwise predictive density} \\
&=& \sum_{i=1}^n \log \left(\frac{1}{S}\sum_{s=1}^S
p(y_i|\theta^s)\right). \label{butgut3}
\end{eqnarray}

\subsection{Leave-one-out cross-validation}
\label{sec:LOO}

The Bayesian LOO estimate of out-of-sample predictive fit is
\begin{equation}
\label{xformula2} \mbox{elpd}_{\rm loo} = \sum_{i=1}^n \log p(y_i|y_{-i}),
\end{equation}
where
\begin{equation}
  p(y_i|y_{-i})=\int p(y_i|\theta)p(\theta|y_{-i}) d\theta
  \label{xformula2b}
\end{equation}
is the leave-one-out predictive density given the data without the $i$th data point.

\paragraph{Raw importance sampling.}
As noted by Gelfand, Dey, and Chang (1992), if the $n$ points are conditionally independent in
the data model we can then evaluate (\ref{xformula2b}) with draws $\theta^s$
from the full posterior $p(\theta|y)$ using importance ratios
\begin{equation}
  \label{LOO-weights}
  r_i^s=\frac{1}{p(y_i|\theta^s)} \propto \frac{p(\theta^s|y_{-i})}{p(\theta^s|y)}
\end{equation}
to get the importance sampling leave-one-out (IS-LOO) predictive distribution,
\begin{align}
  \label{eq:is-loo}
p(\tilde{y}_i|y_{-i})\approx\frac{\sum_{s=1}^S r_i^s p(\tilde{y}_i|\theta^s)}{\sum_{s=1}^S r_i^s}.
\end{align}
Evaluating this LOO log predictive density at the held-out data point $y_i$, we
get
 \begin{equation}
  \label{eq:cpo}
   p(y_i|y_{-i})\approx\frac{1}{\frac{1}{S}\sum_{s=1}^S\frac{1}{p(y_i|\theta^s)}}.
 \end{equation}
However, the posterior $p(\theta|y)$ is likely to have a smaller variance and
thinner tails than the leave-one-out distributions $p(\theta|y_{-i})$, and thus
a direct use of (\ref{eq:cpo}) induces instability because the importance ratios
can have high or infinite variance.

For simple models the variance of the importance weights may be computed
analytically. The necessary and sufficient conditions for the variance of the
case-deletion importance sampling weights to be finite for a Bayesian linear
model are given by Peruggia (1997). Epifani et al.\ (2008) extend the analytical
results to generalized linear models and non-linear Michaelis-Menten models.
However, these conditions can not be computed analytically in general.

Koopman et al.\ (2009) propose to use the maximum likelihood fit of the
generalized Pareto distribution to the upper tail of the distribution of the
importance ratios and use the fitted parameters to form a test for whether the
variance of the importance ratios is finite. If the hypothesis test suggests the
variance is infinite then they abandon importance sampling.

\paragraph{Truncated importance sampling.}

Ionides (2008) proposes a modification of importance sampling where the raw
importance ratios $r^s$ are replaced by truncated weights
\begin{equation}\label{min}
  w^s = \min(r^s,\sqrt{S}\bar{r}),
\end{equation}
where $\bar{r}=\frac{1}{S}\sum_{s=1}^Sr^s$.
Ionides (2008) proves that the variance of the truncated importance sampling
weights is guaranteed to be finite, and provides theoretical and experimental
results showing that truncation using the threshold $\sqrt{S}\bar{r}$ gives an
importance sampling estimate with a mean square error close to an estimate with
a case specific optimal truncation level. The downside of the truncation is that
it introduces a bias, which can be large as we demonstrate in our experiments.

\paragraph{Pareto smoothed importance sampling.}

We can improve the LOO estimate using Pareto smoothed importance sampling (PSIS;
Vehtari and Gelman, 2015), which applies a smoothing procedure to the importance
weights. We briefly review the motivation and steps of PSIS here, before moving
on to focus on the goals of using and evaluating predictive information
criteria.

As noted above, the distribution of the importance weights used in LOO may have
a long right tail. We use the empirical Bayes estimate of Zhang and Stephens
(2009) to fit a generalized Pareto distribution to the tail (20\% largest
importance ratios). By examining the shape parameter $k$ of the fitted Pareto
distribution, we are able to obtain sample based estimates of the existence of
the moments (Koopman et al, 2009). This extends the diagnostic approach of
Peruggia (1997) and Epifani et al.\ (2008) to be used routinely with IS-LOO for
any model with a factorizing likelihood.

Epifani et al.\ (2008) show that when estimating the leave-one-out predictive
density, the central limit theorem holds if the distribution of the weights has
finite variance. These results can be extended via the generalized central limit
theorem for stable distributions. Thus, even if the variance of the importance
weight distribution is infinite, if the mean exists then the accuracy of the
estimate improves as additional posterior draws are obtained.

When the tail of the weight distribution is long, a direct use of importance
sampling is sensitive to one or few largest values. By fitting a generalized
Pareto distribution to the upper tail of the importance weights, we smooth these
values. The procedure goes as follows:
\begin{enumerate}
\item Fit the generalized Pareto distribution to the 20\% largest importance
ratios $r_s$ as computed in (\ref{LOO-weights}).
The computation is done separately for each held-out data point $i$. In
simulation experiments with thousands and tens of thousands of draws, we have
found that the fit is not sensitive to the specific cutoff value (for a
consistent estimation, the proportion of the samples above the cutoff should get
smaller when the number of draws increases).

\item Stabilize the importance ratios by replacing the $M$ largest ratios by the
expected values of the order statistics of the fitted generalized Pareto
distribution
\begin{align*}
  F^{-1}\left(\frac{z-1/2}{M}\right), \quad z=1,\ldots,M,
\end{align*}
where $M$ is the number of simulation draws used to fit the Pareto (in this
case, $M=0.2\,S$) and $F^{-1}$ is the inverse-CDF of the generalized Pareto
distribution. Label these new weights as $\tilde{w}_i^s$ where, again, $s$
indexes the simulation draws and $i$ indexes the data points; thus, for each $i$
there is a distinct vector of $S$ weights.

\item To guarantee finite variance of the estimate, truncate each vector of
weights at $S^{3/4}\bar{w}_i$, where $\bar{w}_i$ is the average of the $S$
smoothed weights corresponding to the distribution holding out data point $i$.
Finally, label these truncated weights as $w^s_i$.
\end{enumerate}
The above steps must be performed for each data point $i$. The result is a
vector of weights $w_i^s, s=1,\dots,S$, for each $i$, which in general should be
better behaved than the raw importance ratios $r_i^s$ from which they are
constructed.

The results can then be combined to compute the desired LOO estimates.
The PSIS estimate of the LOO expected log pointwise predictive density is
\begin{equation}
  \label{LOO}
  \widehat{\mbox{elpd}}_{\rm psis-loo} = \sum_{i=1}^n \log \left(\frac{\sum_{s=1}^S w_i^s p(y_i|\theta^s)}{\sum_{s=1}^S w_i^s}\right).
\end{equation}

The estimated shape parameter $\hat{k}$ of the generalized Pareto distribution
can be used to assess the reliability of the estimate:
\begin{itemize}

\item If $k<\frac{1}{2}$, the variance of the raw importance ratios is finite,
the central limit theorem holds, and the estimate converges quickly.

\item If $k$ is between $\frac{1}{2}$ and 1, the variance of the raw importance
ratios is infinite but the mean exists, the generalized central limit theorem
for stable distributions holds, and the convergence of the estimate is slower.
The variance of the PSIS estimate is finite but may be large.

\item If $k>1$, the variance and the mean of the raw ratios distribution do not
exist. The variance of the PSIS estimate is finite but may be large.

\end{itemize}
If the estimated tail shape parameter $\hat{k}$ exceeds 0.5, the user should be
warned, although in practice we have observed good performance for values of $\hat{k}$ up to 
0.7. Even if the PSIS estimate has a finite variance, when $\hat{k}$ exceeds 0.7 the
user should consider sampling directly from $p(\theta^s|y_{-i})$ for the
problematic $i$, use $K$-fold cross-validation (see Section~\ref{sec:K-fold-CV}), or use a more robust model.

The additional computational cost of sampling directly from each
$p(\theta^s|y_{-i})$ is approximately the same as sampling from the full
posterior, but it is recommended if the number of problematic data points is not too high.

A more robust model may also help because importance sampling is less likely
to work well if the marginal posterior $p(\theta^s|y)$ and LOO posterior
$p(\theta^s|y_{-i})$ are very different. This is more likely to happen with a
non-robust model and highly influential observations. A robust model may reduce
the sensitivity to one or several highly influential observations, as we show in
the examples in Section \ref{sec:examples}.

\subsection{WAIC}
\label{sec:waic}

WAIC (Watanabe, 2010) is an alternative approach to estimating the expected log
pointwise predictive density and is defined as
\begin{equation}
\label{eq:WAIC}
\widehat{\mbox{elpd}}_{\rm waic} = \widehat{\mbox{lpd}} - \widehat{p}_{\rm waic},
\end{equation}
where $\widehat{p}_{\rm waic}$ is the estimated effective number of parameters
and is computed based on the definition\footnote{In Gelman, Carlin, et al.\ (2013),
the variance-based $p_{\rm waic}$ defined here is called $p_{{\rm waic}\, 2}$.
There is also a mean-based formula, $p_{{\rm waic}\, 1}$, which we do not use
here.}
\begin{equation}\label{cp0}
  p_{\rm waic} =\sum_{i=1}^n \mbox{var}_{\rm post} \left(\log p(y_i|\theta)\right),
\end{equation}
which we can calculate using the posterior variance of the log predictive
density for each data point $y_i$, that is, $V_{s=1}^S \log p(y_i|\theta^s)$,
where $V_{s=1}^S$ represents the sample variance,
$V_{s=1}^S a_s = \frac{1}{S-1}\sum_{s=1}^S (a_s - \bar{a})^2$. Summing over all
the data points $y_i$ gives a simulation-estimated effective number of
parameters,
\begin{equation}\label{cp}
 \widehat{p}_{\rm waic} = \sum_{i=1}^n V_{s=1}^S\left( \log p(y_i|\theta^s)\right).
\end{equation}

For DIC, there is a similar variance-based computation of the number of
parameters that is notoriously unreliable, but the WAIC version is more stable
because it computes the variance separately for each data point and then takes
the sum; the summing yields stability.

The {\em effective number of parameters} $\widehat{p}_{\rm waic}$ can be
used as measure of complexity of the model, but it should not be overinterpreted,
as the original goal is to estimate the difference between lpd and elpd.
As shown by Gelman, Hwang, and Vehtari (2014) and demonstrated also in Section
\ref{sec:examples}, in the case of a weak prior, $\widehat{p}_{\rm waic}$ can
severely underestimate the difference between lpd and elpd.
For $\widehat{p}_{\rm waic}$ there is no similar theory as for the moments of
the importance sampling weight distribution, but based on our simulation
experiments it seems that $\widehat{p}_{\rm waic}$ is unreliable if any of the
terms $V_{s=1}^S \log p(y_i|\theta^s)$ exceeds 0.4.

The different behavior of LOO and WAIC seen in the experiments can be
understood by comparing Taylor series approximations.
By defining a generating function of functional cumulants,
\begin{align}
F(\alpha)=\sum_{i=1}^n\log E_{\rm post}(p(y_i|\theta)^\alpha),
\end{align}
and applying a Taylor expansion of $F(\alpha)$ around 0 with $\alpha=-1$ we
obtain an expansion of $\mbox{lpd}_{\rm loo}$
\begin{align}
 \mbox{elpd}_{\rm loo}=F'(0)-\frac{1}{2}F''(0)+\frac{1}{6}F^{(3)}(0)-\sum_{i=4}^\infty\frac{(-1)^iF^{(i)}(0)}{i!}.
\end{align}
From the definition of $F(\alpha)$ we get
\begin{align}
  F(0)&=0\nonumber \\
  F(1)&=\sum_{i=1}^n\log E_{\rm post} (p(y_i|\theta))\nonumber \\
  F'(0)&=\sum_{i=1}^n E_{\rm post} ( \log p(y_i|\theta))\nonumber \\
  F''(0)&=\sum_{i=1}^n \mbox{var}_{\rm post}(\log p(y_i|\theta)).
\end{align}
Furthermore
\begin{align}
  \mbox{lpd}=F(1)=F'(0)+\frac{1}{2}F''(0)+\frac{1}{6}F^{(3)}(0)+\sum_{i=4}^\infty\frac{F^{(i)}(0)}{i!},
\end{align}
and the expansion for WAIC is then
\begin{align}
\text{WAIC}&=F(1)-F''(0)\nonumber\\
&=F'(0)-\frac{1}{2}F''(0)+\frac{1}{6}F^{(3)}(0)+\sum_{i=4}^\infty\frac{F^{(i)}(0)}{i!}.
\end{align}
The first three terms of the expansion of WAIC match the expansion of LOO, and
the rest of the terms match the expansion of lpd. Watanabe (2010) argues that,
asymptotically, the latter terms have negligible contribution and thus
asymptotic equivalence with LOO is obtained. However, the error can be
significant in the case of finite $n$ and weak prior information as shown by
Gelman, Hwang, and Vehtari (2014), and demonstrated also in Section
\ref{sec:examples}.
If the higher order terms are not negligible, then WAIC is biased towards lpd.
To reduce this bias it is possible to compute additional series terms, but
computing higher moments using a finite posterior sample increases the variance
of the estimate and, based on our experiments, it is more difficult to control
the bias-variance tradeoff than in PSIS-LOO.
WAIC's larger bias compared to LOO is also demonstrated by Vehtari et al.\ (2016) 
in the case of Gaussian processes with distributional posterior approximations.
In the experiments we also demonstrate the we can use truncated IS-LOO with
heavy truncation to obtain similar bias towards lpd and similar estimate
variance as in WAIC.

\subsection{$K$-fold cross-validation}
\label{sec:K-fold-CV}

In this paper we focus on leave-one-out cross-validation and WAIC, but, for
statistical and computational reasons, it can make sense to cross-validate using
$K <\!<n$ hold-out sets.  In some ways, $K$-fold cross-validation is simpler than
leave-one-out cross-validation but in other ways it is not.  $K$-fold
cross-validation requires refitting the model $K$ times which can be
computationally expensive whereas approximative LOO methods, such as PSIS-LOO,
require only one evaluation of the model.

If in PSIS-LOO $\hat{k}>0.7$ for a few $i$ we recommend sampling directly from
each corresponding $p(\theta^s|y_{-i})$, but if there are more than $K$
problematic $i$, then we recommend checking the results using $K$-fold
cross-validation. Vehtari \& Lampinen (2002) demonstrate cases where IS-LOO
fails (according to effective sample size estimates instead of the $\hat{k}$
diagnostic proposed here) for a large number of $i$ and K-fold-CV produces more
reliable results.

In Bayesian $K$-fold cross-validation, the data are partitioned into $K$ subsets
$y_k$, for $k=1,\dots,K$, and then the model is fit separately to each training
set $y_{(-k)}$, thus yielding a posterior distribution
$p_{{\rm post} (-k)}(\theta)\!=\!p(\theta|y_{(-k)})$.
If the number of partitions is small (a typical value in the literature is
$K=10$), it is not so costly to simply re-fit the model separately to each
training set. To maintain consistency with LOO and WAIC, we define predictive
accuracy for each data point, so that the log predictive density for $y_i$, if
it is in subset $k$, is
\begin{equation}
\log p(y_i|y_{(-k)})=\log\!\int\! \!p(y_i|\theta)p(\theta|y_{(-k)})d\theta, \quad i \in k.
\end{equation}
Assuming the posterior distribution $p(\theta|y_{(-k)})$ is summarized by $S$
simulation draws $\theta^{k,s}$, we calculate its log predictive density as
\begin{equation}
\widehat{{\rm elpd}}_i = \log\left(\frac{1}{S}\sum_{s=1}^S p(y_i|\theta^{k,s})\right)
\end{equation}
using the simulations corresponding to the subset $k$ that contains data point
$i$.  We then sum to get the estimate
\begin{equation}
\label{holdout}
\widehat{\rm elpd}_{\rm xval}=\sum_{i=1}^n \widehat{\rm elpd}_i.
\end{equation}
There remains a bias as the model is learning from a fraction $\frac{1}{K}$ less of the data. 
Methods for correcting this bias exist but are rarely used as they can increase the variance, 
and if $K \geq 10$ the size of the bias is typically small compared to the variance of the 
estimate (Vehtari and Lampinen, 2002).
In our experiments, exact LOO is the same as $K$-fold-CV with $K=N$ and we also
analyze the effect of this bias and bias correction in
Section~\ref{sec:sim8schools}.

For $K$-fold cross-validation, if the subjects are exchangeable, that is, the
order does not contain information, then there is no need for random selection.
If the order does contain information, e.g. in survival studies the later
patients have shorter follow-ups, then randomization is often useful.

In most cases we recommend partitioning the data into subsets by randomly
permuting the observations and then systemically dividing them into $K$
subgroups. 
If the subjects are exchangeable, that is, the order does not contain
information, then there is no need for random selection, but if the order does
contain information, e.g. in survival studies the later patients have shorter
follow-ups, then randomization is useful.
In some cases it may be useful to stratify to obtain better balance among
groups. See Vehtari and Lampinen (2002), Celisse and Arlot (2010), and Vehtari
and Ojanen (2012) for further discussion of these points.

As the data can be divided in many ways into $K$ groups it introduces
additional variance in the estimates, which is also evident from our
experiments. This variance can be reduced by repeating $K$-fold-CV
several times with different permutations in the data division, but
this will further increase the computational cost.

\subsection{Data division}

The purpose of using LOO or WAIC is to estimate the accuracy of the predictive
distribution $p(\tilde{y}_i|y)$. Computation of PSIS-LOO and WAIC (and AIC and
DIC) is based on computing terms
$\log p(y_i|y)=\log\!\int \!p(y_i|\theta)p(\theta|y)$
assuming some agreed-upon division of the data $y$ into individual data points
$y_i$. Although often $y_i$ will denote a single scalar observation, in the case
of hierarchical data, it may denote a group of observations. For example, in
cognitive or medical studies we may be interested in prediction for a new
subject (or patient), and thus it is natural in cross-validation to consider an
approach where $y_i$ would denote all observations for a single subject and
$y_{-i}$ would denote the observations for all the other subjects. In theory, we
can use PSIS-LOO and WAIC in this case, too, but as the number of observations
per subject increases it is more likely that they will not work as well. The
fact that importance sampling is difficult in higher dimensions is well known
and is demonstrated for IS-LOO by Vehtari and Lampinen (2002) and for PSIS by
Vehtari and Gelman (2015). The same problem can also be shown to hold
for WAIC. If diagnostics warn about the reliability of PSIS-LOO (or WAIC), then
$K$-fold cross-validation can be used by taking into account the hierarchical
structure in the data when doing the data division as demonstrated, for example,
by Vehtari and Lampinen (2002).

\section{Implementation in Stan}
\label{sec:stan}

We have set up code to implement LOO, WAIC, and $K$-fold cross-validation in R
and Stan so that users will have a quick and convenient way to assess and
compare model fits.  Implementation is not automatic, though, because of the
need to compute the separate factors $p(y_i|\theta)$ in the likelihood. Stan
works with the joint density and in its usual computations does not ``know''
which parts come from the prior and which from the likelihood. Nor does Stan in
general make use of any factorization of the likelihood into pieces
corresponding to each data point.  Thus, to compute these measures of predictive
fit in Stan, the user needs to explicitly code the factors of the likelihood
(actually, the terms of the log-likelihood) as a vector.  We can then pull apart
the separate terms and compute cross-validation and WAIC at the end, after all
simulations have been collected. Sample code for carrying out this procedure
using Stan and the {\tt loo} R package is provided in Appendix~\ref{sec:rstanarm}.  This code can
be adapted to apply our procedure in other computing languages.

Although the implementation is not automatic when writing custom Stan programs, 
we can create implementations that are automatic for users of our new {\tt rstanarm} 
R package (Gabry and Goodrich, 2016). {\tt rstanarm} provides a high-level interface to 
Stan that enables the user to specify many of the most common applied Bayesian regression 
models using standard R modeling syntax (e.g. like that of {\tt glm}). 
The models are then estimated using Stan's algorithms and the results are returned to the 
user in a form similar to the fitted model objects to which R users are accustomed. 
For the models implemented in {\tt rstanarm}, we have preprogrammed many
tasks, including computing and saving the pointwise predictive measures and importance ratios which we use to compute WAIC and PSIS-LOO. The {\tt loo} method for {\tt rstanarm} models requires no 
additional programming from the user after fitting a model, as we can compute 
all of the needed quantities internally from the contents of the fitted model object
and then pass them to the functions in the {\tt loo} package. Examples of using 
{\tt loo} with {\tt rstanarm} can be found in the {\tt rstanarm} vignettes, and we 
also provide an example in Appendix~\ref{sec:rstanarm} of this paper. 

\section{Examples}
\label{sec:examples}

We illustrate with six simple examples:  two examples from our earlier research
in computing the effective number of parameters in a hierarchical model, three
examples that were used by Epifani et al.\ (2008) to illustrate the estimation
of the variance of the weight distribution, and one example of a multilevel
regression from our earlier applied research. For each example we used the Stan
default of 4 chains run for 1000 warmup and 1000 post-warmup iterations,
yielding a total of 4000 saved simulation draws.  With Gibbs sampling or
random-walk Metropolis, 4000 is not a large number of simulation draws.  The
algorithm used by Stan is Hamiltonian Monte Carlo with No-U-Turn-Sampling
(Hoffman and Gelman, 2014), which is much more efficient, and 1000 is already
more than sufficient in many real-world settings.  In these examples we followed
standard practice and monitored convergence and effective sample sizes as
recommended by Gelman, Carlin, et al.\ (2013). We performed 100 independent 
replications of all experiments to obtain estimates of variation. For the exact LOO 
results and convergence plots we run longer chains to obtain a total of 100,000 draws 
(except for the radon example which is much slower to run).

\subsection{Example:  Scaled 8 schools}\label{schools}

\begin{table}
\small{
\begin{center}\begin{tabular}{ccc}
&\multicolumn{1}{c}{Estimated}&\multicolumn{1}{c}{Standard error}\\
School&\multicolumn{1}{c}{effect, $y_j$}&
\multicolumn{1}{c}{of estimate, $\sigma_j$}\\\hline
A& \ 28 & 15 \\
B& \ \,\, 8 & 10 \\
C& $\,-3$ & 16 \\
D& \ \,\, 7 & 11 \\
E& $\,-1$ & \ 9 \\
F& \ \,\, 1 & 11 \\
G& \ 18 & 10 \\
H& \ 12 & 18
\end{tabular}\end{center}
}
\vspace{-.2in}
\caption{\em In a controlled study, independent randomized experiments were
conducted in 8 different high schools to estimate the effect of special
preparation for college admission tests. Each row of the table gives an estimate
and standard error from one of the schools. A hierarchical Bayesian model was
fit to perform meta-analysis and use partial pooling to get more accurate
estimates of the 8 effects. From Rubin (1981).}\label{tab5.2}
\end{table}

\begin{figure}
\center{
   \includegraphics[width=.285\textwidth]{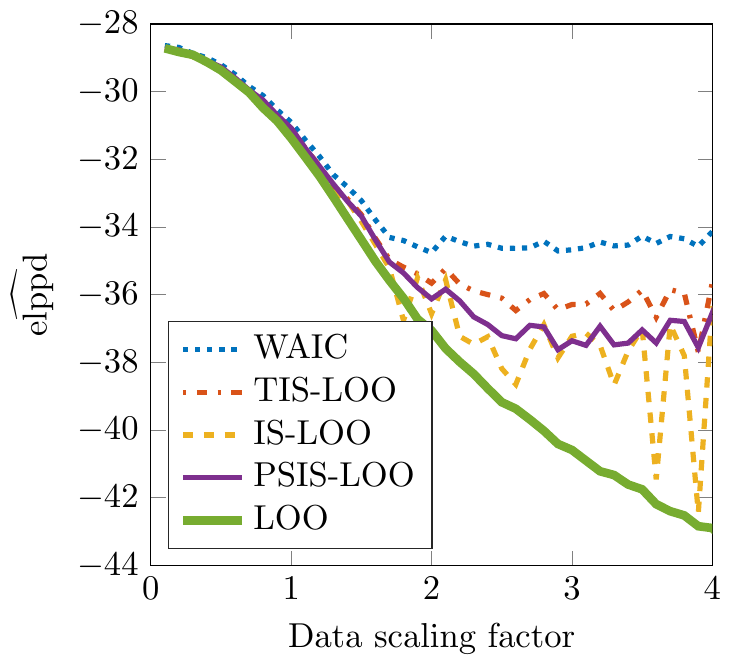}\hspace{.5cm}
  \includegraphics[width=.275\textwidth]{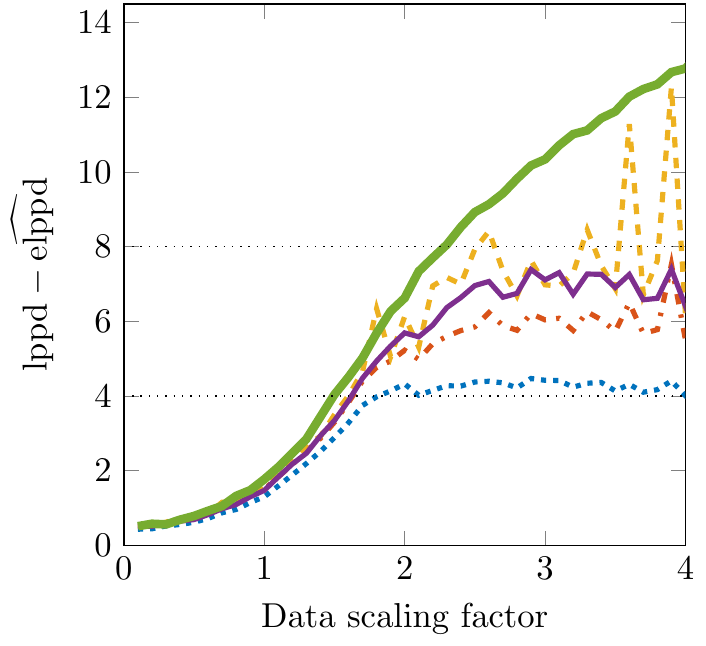}\\
  \includegraphics[width=.28\textwidth]{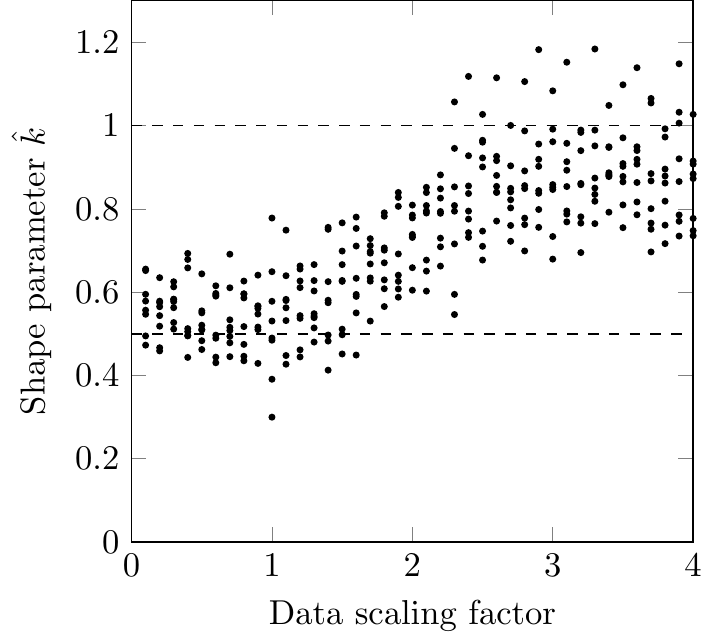}\hspace{.5cm}
  \includegraphics[width=.28\textwidth]{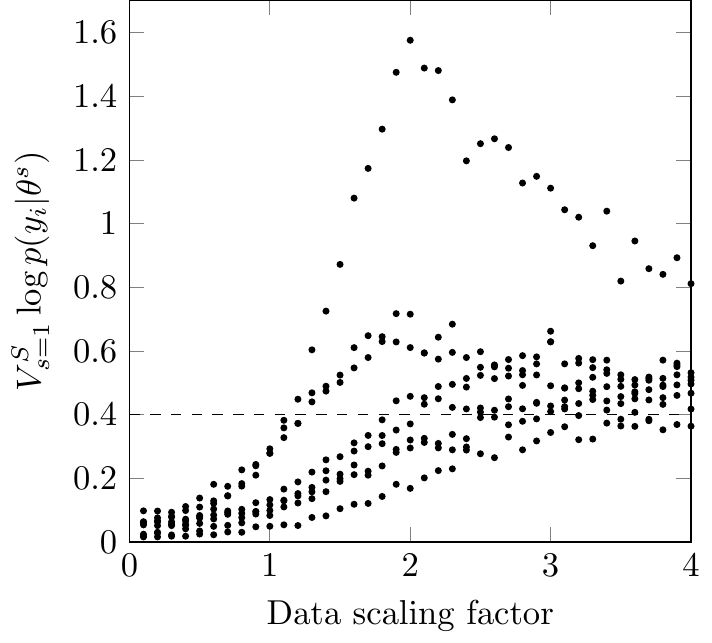}
}
\caption{\em 8 schools example:  (a) WAIC, Truncated Importance Sampling LOO,
Importance Sampling LOO, Pareto Smoothed Importance Sampling LOO, and exact LOO
(which in this case corresponds to 8-fold-CV); (b) estimated effective number of
parameters for each of these measures; (c) tail shape $\hat{k}$ for the importance
weights; and (d) the posterior variances of the log predictive densities, for
scaled versions of the 8 schools data (the original observations $y$ have been
multiplied by a common factor). We consider scaling factors ranging from 0.1
(corresponding to near-zero variation of the underlying parameters among the
schools) to 4 (implying that the true effects in the schools vary by much more
than their standard errors of measurement).  As the scaling increases,
eventually LOO approximations and WAIC fail to approximate exact LOO as the
leave-one-out posteriors are not close to the full posterior. When the estimated
tail shape $\hat{k}$ exceeds 1, the importance-weighted LOO approximations start to
fail. When posterior variances of the log predictive densities exceeds 0.4, WAIC
starts to fail.  PSIS-LOO performs the best among the approximations considered
here.}
\label{fig:scaled8schools}
\end{figure}

For our first example we take an analysis of an education experiment used by
Gelman, Hwang, and Vehtari (2014) to demonstrate the use of information criteria
for hierarchical Bayesian models.

The goal of the study was to measure the effects of a test preparation program
conducted in eight different high schools in New Jersey.  A separate randomized
experiment was conducted in each school, and the administrators of each school
implemented the program in their own way.  Rubin (1981) performed a Bayesian
meta-analysis, partially pooling the eight estimates toward a common mean. The
model has the form, $y_i\sim {\rm N}(\theta_i,\sigma^2_i)$ and $\theta_i\sim
{\rm N}(\mu,\tau^2)$, for $i=1,\dots,n=8$, with a uniform prior distribution on
$(\mu,\tau)$. The measurements $y_i$ and uncertainties $\sigma_i$ are the
estimates and standard errors from separate regressions performed for each
school, as shown in Table \ref{tab5.2}.  The test scores for the individual
students are no longer available.

This model has eight parameters but they are constrained through their
hierarchical distribution and are not estimated independently; thus we would
anticipate the effective number of parameters should be some number between 1
and 8.

To better illustrate the behavior of LOO and WAIC, we repeat the analysis,
rescaling the data points $y$ by a factor ranging from 0.1 to 4 while keeping
the standard errors $\sigma$ unchanged. With a small data scaling factor the
hierarchical model nears complete pooling and with a large data scaling factor
the model approaches separate fits to the data for each school. Figure
\ref{fig:scaled8schools} shows $\widehat{\mbox{elpd}}$ for the various LOO
approximation methods as a function of the scaling factor, based on 4000
simulation draws at each grid point.

When the data scaling factor is small (here, less than 1.5), all
measures largely agree. As the data scaling factor increases and the model
approaches no pooling, the population prior for $\theta_i$ gets flat and
$p_{\rm waic}\approx \frac{p}{2}$. This is correct behavior, as discussed by
Gelman, Hwang, and Vehtari (2014).

\begin{figure}
\centerline{
   \includegraphics[width=.33\textwidth]{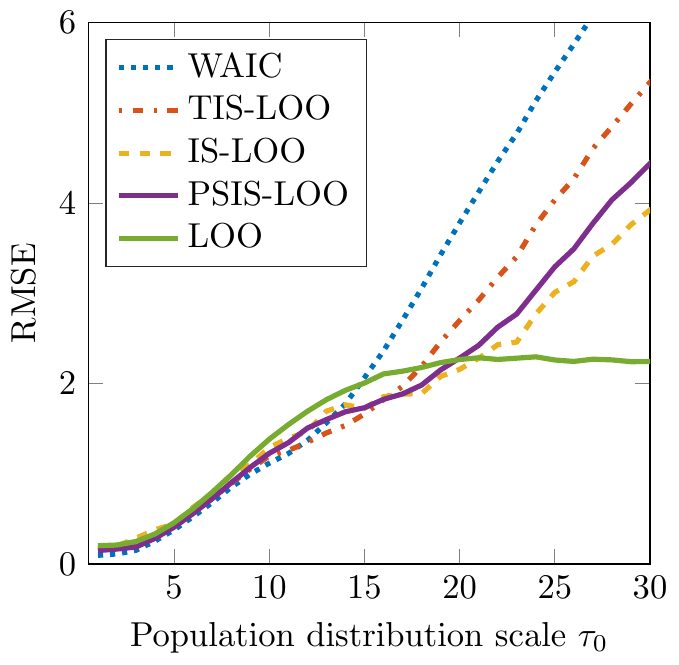}
   \includegraphics[width=.33\textwidth]{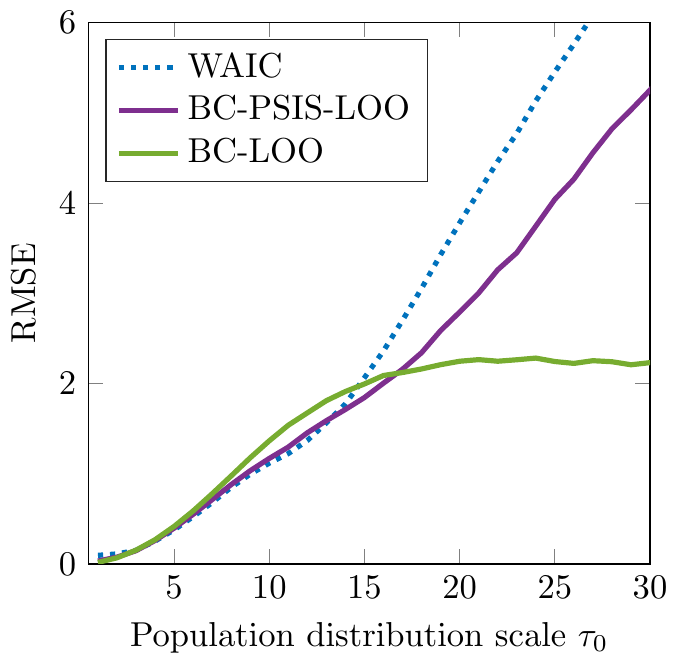}
   \includegraphics[width=.33\textwidth]{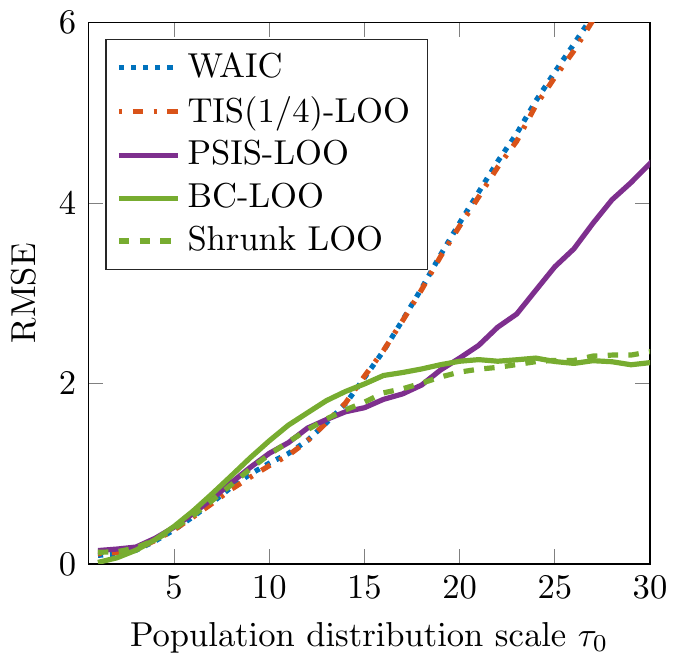}
}
\caption{\em Simulated 8 schools example:  (a) Root mean square error of WAIC,
Truncated Importance Sampling LOO, Importance Sampling LOO, Pareto Smoothed
Importance Sampling LOO, and exact LOO with the true predictive performance
computed using independently simulated test data; the error for all the methods
increases, but the RMSE of exact LOO has an upper limit.  Eventually the LOO
approximations and WAIC fail to return exact LOO, as the leave-one-out
posteriors are not close to the full posterior. When the estimated tail shape
$k$ exceeds 1, the importance-weighted LOO approximations start to fail.  Among
the approximations IS-LOO has the smallest RMSE as it has the smallest bias, and
as the tail shape $k$ is mostly below 1, it does not fail badly.\newline (b)
Root mean square error of WAIC, bias corrected Pareto Smoothed Importance
Sampling LOO, and bias corrected exact LOO with the true predictive performance
computed using independently simulated test data. The bias correction also
reduces RMSE, having the clearest impact with smaller population distribution
scales, but overall the reduction in RMSE is negligible.\newline (c) Root mean
square error of WAIC, Truncated Importance Sampling LOO with heavy truncation
($\sqrt[4]{S}\bar{r}$), Pareto Smoothed Importance Sampling LOO, bias corrected
exact LOO, and shrunk exact LOO with the true predictive performance computed
using independently simulated test data. Truncated Importance Sampling LOO with
heavy truncation matches WAIC accurately. Shrinking exact LOO towards
the lpd of observed data reduces the RMSE for some scale values with small
increase in error for larger scale values.}
\label{fig:sim8schools1}
\end{figure}

In the case of exact LOO, $\widehat{\rm lpd} - \widehat{\rm elpd}_{\rm loo}$ can
be larger than $p$. As the prior for $\theta_i$ approaches flatness, the log
predictive density $p_{{\rm post}(-i)}(y_i) \rightarrow -\infty$. At the same
time, the full posterior becomes an inadequate approximation to
$p_{{\rm post}(-i)}$ and all approximations become poor approximations to the
actual out-of-sample prediction error under the model.

WAIC starts to fail when one of the posterior variances of the log predictive
densities exceeds 0.4. LOO approximations work well even if the tail shape $k$
of the generalized Pareto distribution is between $\frac{1}{2}$ and 1, and the
variance of the raw importance ratios is infinite. The error of LOO
approximations increases with $k$, with a clearer difference between the methods
when $k>0.7$.

\subsection{Example:  Simulated 8 schools}
\label{sec:sim8schools}

In the previous example, we used exact LOO as the gold standard. In this
section, we generate simulated data from the same statistical model and compare
predictive performance on independent test data. Even when the number of
observations $n$ is fixed, as the scale of the population distribution increases
we observe the effect of weak prior information in hierarchical models discussed
in the previous section and by Gelman, Hwang, and Vehtari (2014). Comparing the
error, bias and variance of the various approximations, we find that PSIS-LOO
offers the best balance.

For $i=1,\dots,n=8$, we simulate $\theta_{0,i}\sim {\rm N}(\mu_0,\tau^2_0)$ and
$y_i\sim {\rm N}(\theta_{0,i},\sigma^2_{0,i})$, where we set $\sigma_{0,i}=10$,
$\mu_0=0$, and $\tau_0\in \{1,2,\ldots,30\}$. The simulated data is similar to
the real 8 schools data, for which the empirical estimate is $\hat{\tau}\approx
10$.  For each value of $\tau_0$ we generate 100 training sets of size 8 and one
test data set of size 1000. Posterior inference is based on 4000 draws for each
constructed model.

Figure \ref{fig:sim8schools1}a shows the root mean square error (RMSE) for the
various LOO approximation methods as a function of $\tau_0$, the scale of the
population distribution. When $\tau_0$ is large all of the approximations
eventually have ever increasing RMSE, while exact LOO has an upper limit. For
medium scales the approximations have {\it smaller} RMSE than exact LOO. As
discussed later, this is explained by the difference in the variance of the
estimates. For small scales WAIC has slightly smaller RMSE than the other
methods (including exact LOO).

Watanabe (2010) shows that WAIC gives an asymptotically unbiased estimate of the
out-of-sample prediction error---this does {\it not} hold for hierarchical
models with weak prior information as shown by Gelman, Hwang, and Vehtari
(2014)---but exact LOO is slightly biased as the LOO posteriors use only $n-1$
observations. WAIC's different behavior can be understood through the truncated
Taylor series correction to the lpd, that is, not using the entire series will
bias it towards lpd (see Section \ref{sec:waic}). The bias in LOO is negligible
when $n$ is large, but with small $n$ it can be be larger.

Figure \ref{fig:sim8schools1}b shows RMSE for the bias corrected LOO
approximations using the first order correction of Burman (1989). For small
scales the error of bias corrected LOOs is smaller than WAIC. When the scale
increases the RMSEs are close to the non-corrected versions. Although the bias
correction is easy to compute, the difference in accuracy is negligible for most
applications.

\begin{figure}
\centerline{
   \includegraphics[width=.33\textwidth]{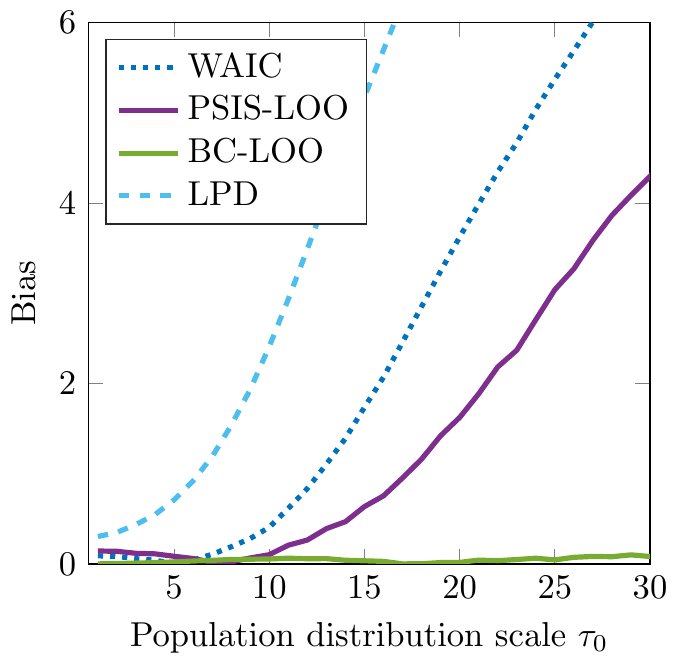}\hspace{.5cm}
  \includegraphics[width=.345\textwidth]{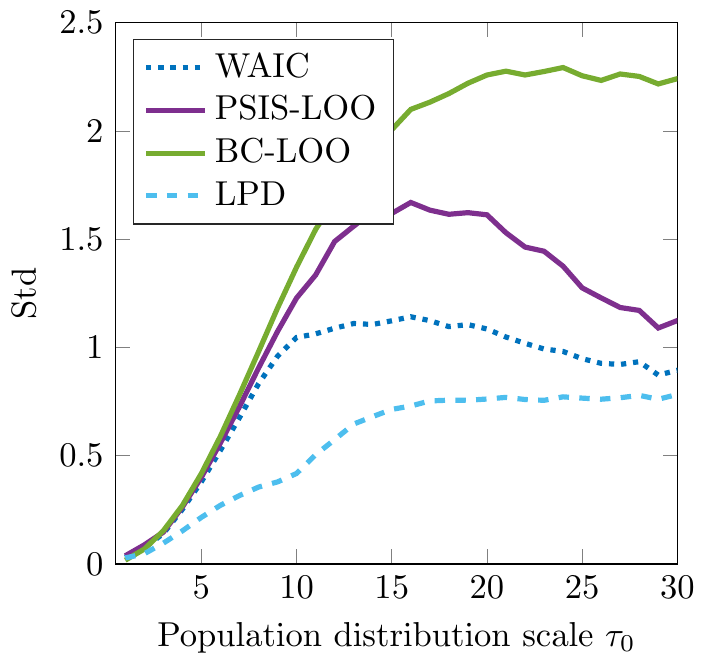}
}
\caption{\em Simulated 8 schools example:  (a) Absolute bias of WAIC, Pareto
Smoothed Importance Sampling LOO, bias corrected exact LOO, and the lpd (log
predictive density) of observed data with the true predictive performance
computed using independently simulated test data; (b) standard deviation for
each of these measures; All methods except the lpd of observed data have small
biases and variances with small population distribution scales. When the scale
increases the bias of WAIC increases faster than the bias of the other methods
(except the lpd of observed data). Bias corrected exact LOO has practically zero
bias for all scale values. WAIC and Pareto Smoothed Importance Sampling LOO have
lower variance than exact LOO, as they are shrunk towards the lpd of observed
data, which has the smallest variance with all scales.}
\label{fig:sim8schools3}
\end{figure}

We shall discuss Figure \ref{fig:sim8schools1}c in a moment, but first consider
Figure \ref{fig:sim8schools3}, which shows the RMSE of the approximation methods
and the lpd of observed data decomposed into bias and standard deviation.
All methods (except the lpd of observed data) have small biases and variances
with small population distribution scales.
Bias corrected exact LOO has practically zero bias for all scale values but the
highest variance. When the scale increases the LOO approximations eventually
fail and bias increases. As the approximations start to fail, there is a certain
region where implicit shrinkage towards the lpd of observed data decelerates the
increase in RMSE as the variance is reduced, even if the bias continues to grow.

If the goal were to minimize the RMSE for smaller and medium scales, we could
also shrink exact LOO and increase shrinkage in approximations. Figure
\ref{fig:sim8schools1}c shows the RMSE of the LOO approximations with two new
choices. Truncated Importance Sampling LOO with very heavy truncation (to
$\sqrt[4]{S}\bar{r}$) closely matches the performance of WAIC. In the
experiments not included here, we also observed that adding more correct Taylor
series terms to WAIC will make it behave similar to Truncated Importance
Sampling with less truncation (see discussion of Taylor series expansion in
Section \ref{sec:waic}). Shrunk exact LOO ($\alpha \cdot \mbox{elpd}_{\rm loo} +
(1-\alpha)\cdot \mbox{lpd}$, with $\alpha=0.85$ chosen by hand for illustrative
purposes only) has a smaller RMSE for small and medium scale values as the
variance is reduced, but the price is increased bias at larger scale values.

If the goal is robust estimation of predictive performance, then exact LOO is
the best general choice because the error is limited even in the case of weak
priors. Of the approximations, PSIS-LOO offers the best balance as well as
diagnostics for identifying when it is likely failing.

\subsection{Example:  Linear regression for stack loss data}

To check the performance of the proposed diagnostic for our second
example we analyze the stack loss data used by Peruggia (1997) which
is known to have analytically proven infinite variance of one of the
importance weight distributions.

The data consist of $n = 21$ daily observations on one outcome and three
predictors pertaining to a plant for the oxidation of ammonia to nitric acid.
The outcome $y$ is an inverse measure of the efficiency of the plant and the
three predictors $x_1$, $x_2$, and $x_3$ measure rate of operation, temperature
of cooling water, and (a transformation of the) concentration of circulating
acid.

\begin{figure}
\centerline{
   \includegraphics[width=.45\textwidth]{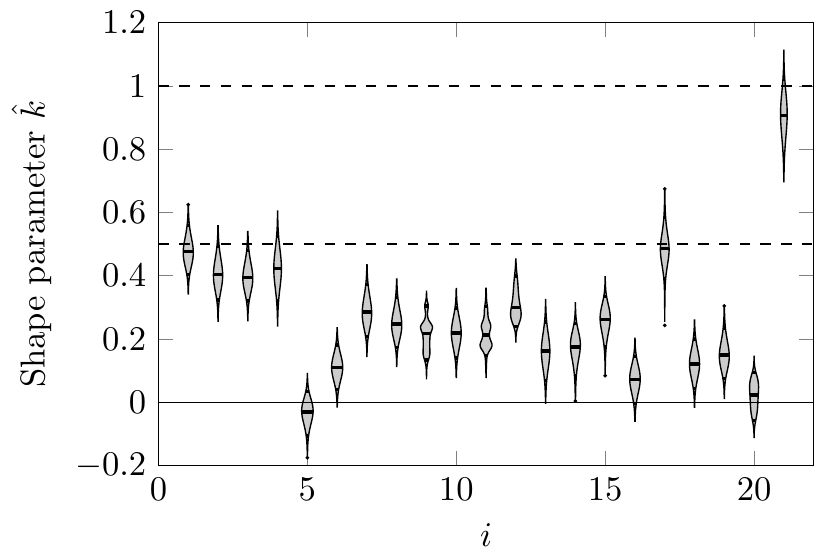}\hspace{.5cm}
  \includegraphics[width=.45\textwidth]{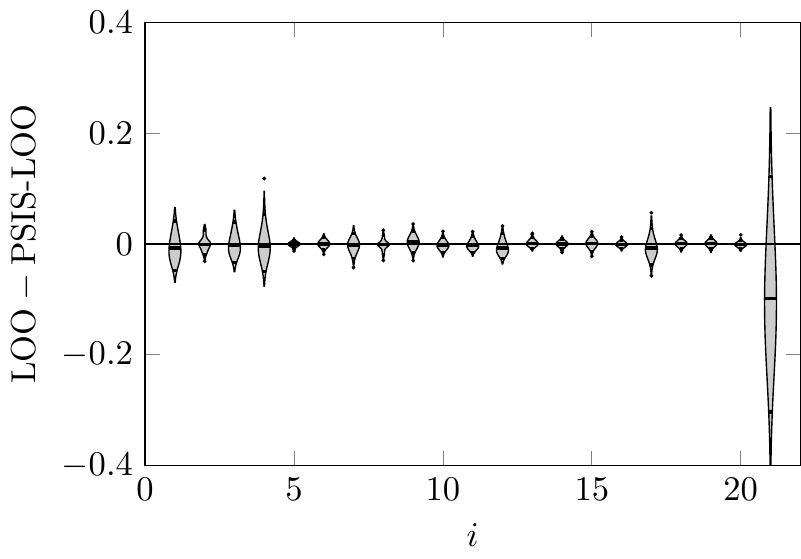}
}
\caption{\em Stack loss example with normal errors:  Distributions of (a) tail
shape estimates and (b) PSIS-LOO estimation errors compared to LOO, from 100
independent Stan runs.  The pointwise calculation of the terms in PSIS-LOO
reveals that much of the uncertainty comes from a single data point, and it
could make sense to simply re-fit the model to the subset and compute LOO
directly for that point.}
\label{fig:stacks_N}
\end{figure}

Peruggia (1997) shows that the importance weights for leave-one-out
cross-validation for the data point $y_{21}$ have infinite variance. Figure
\ref{fig:stacks_N} shows the distribution of the estimated tail shapes $k$ and
estimation errors compared to LOO in 100 independent Stan
runs.\footnote{Smoothed density estimates were made using a logistic Gaussian
process (Vehtari and Riihim{\"a}ki, 2014).} The estimates of the tail shape $k$ for $i=21$ 
suggest that the variance of the raw importance ratios is infinite, however the generalized 
central limit theorem for stable distributions holds and we can still obtain an accurate 
estimate of the component of LOO for this data point using PSIS.

\begin{figure}
\centerline{
   \includegraphics[width=.45\textwidth]{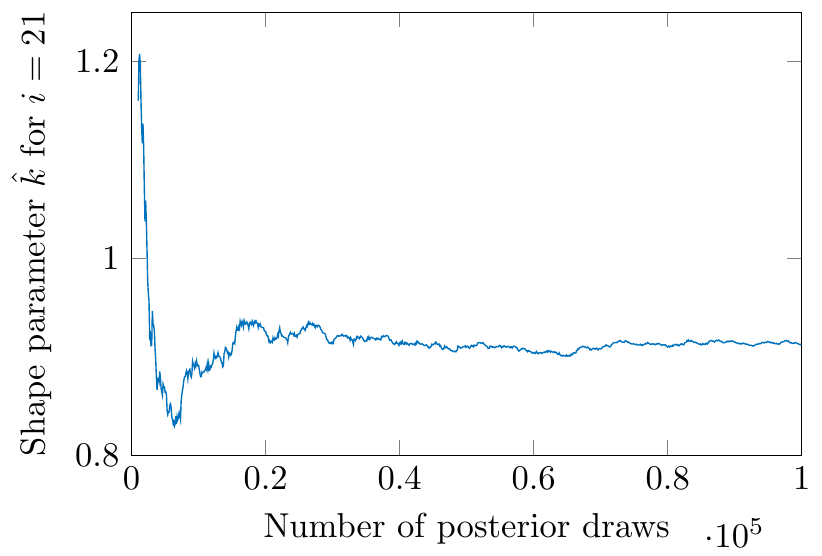}\hspace{.5cm}
  \includegraphics[width=.47\textwidth]{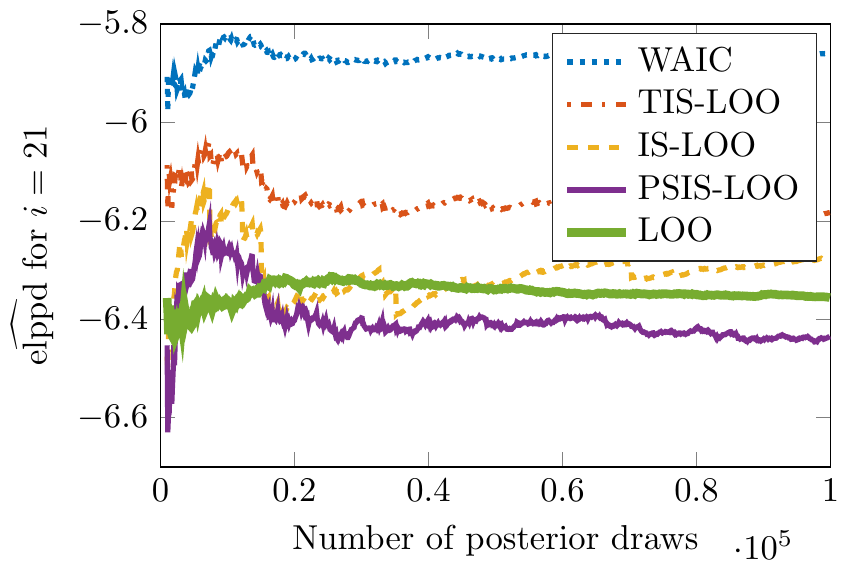}
}
\caption{\em Stack loss example with normal errors:  (a) Tail shape estimate and
(b) LOO approximations for the difficult point, $i=21$. When more draws are
obtained, the estimates converge (slowly) following the generalized central
limit theorem.}
\label{fig:stacks_21}
\end{figure}

Figure \ref{fig:stacks_21} shows that if we continue sampling, the estimates for
both the tail shape $k$ and $\widehat{\rm elpd}_i$ do converge (although slowly
as $k$ is close to 1). As the convergence is slow it would be more efficient to
sample directly from $p(\theta^s|y_{-i})$ for the problematic $i$.

\begin{figure}
\centerline{
   \includegraphics[width=.45\textwidth]{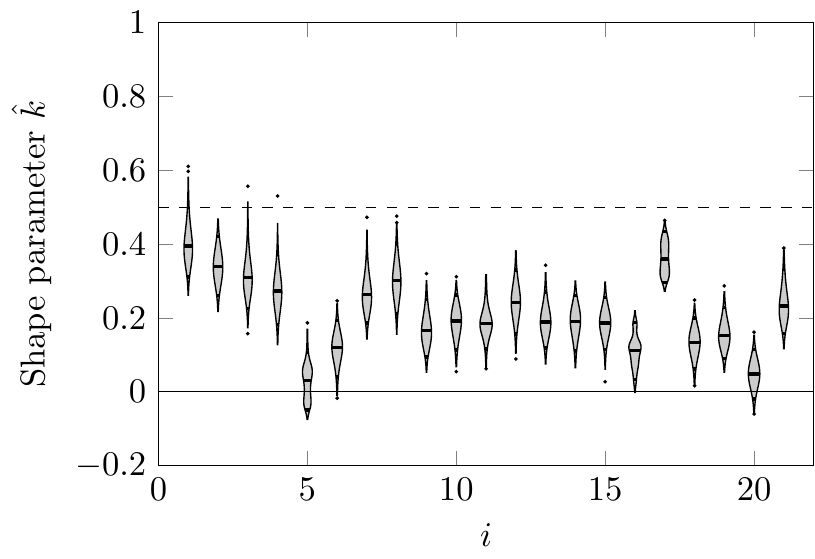}\hspace{.5cm}
  \includegraphics[width=.45\textwidth]{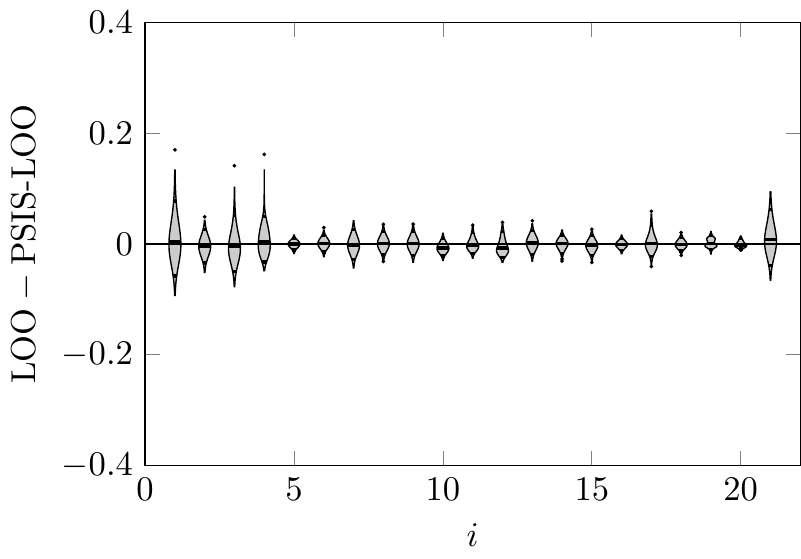}
}
\caption{\em Stack loss example with Student-$t$ errors:   Distributions of (a)
tail shape estimates and (b) PSIS-LOO estimation errors compared to LOO, from
100 independent Stan runs.  The computations are more stable than with normal
errors (compare to Figure \ref{fig:stacks_N}).}
\label{fig:stacks_t}
\end{figure}

High estimates of the tail shape parameter $k$ indicate that the full posterior
is not a good importance sampling approximation to the desired leave-one-out
posterior, and thus the observation is surprising according to the model. It is
natural to consider an alternative model.  We tried replacing the normal
observation model with a Student-$t$ to make the model more robust for the
possible outlier.  Figure \ref{fig:stacks_t} shows the distribution of the
estimated tail shapes $\hat{k}$ and estimation errors for PSIS-LOO compared to
LOO in 100 independent Stan runs for the Student-$t$ linear regression model.
The estimated tail shapes and the errors in computing this component of LOO are
smaller than with Gaussian model.

\subsection{Example: Nonlinear regression for Puromycin reaction data}

As a nonlinear regression example, we use the Puromycin biochemical reaction
data also analyzed by Epifani et al.\ (2008).
For a group of cells not treated with the drug Puromycin, there are $n = 11$
measurements of the initial velocity of a reaction, $V_i$ , obtained when the
concentration of the substrate was set at a given positive value, $c_i$.
Velocity on concentration is given by the Michaelis-Menten relation,
$V_i \sim \mbox{N}(mc_i/(\kappa + c_i), \sigma^2)$. Epifani et al.\ (2008) show
that the raw importance ratios for observation $i=1$ have infinite variance.

\begin{figure}
\centerline{
   \includegraphics[width=.45\textwidth]{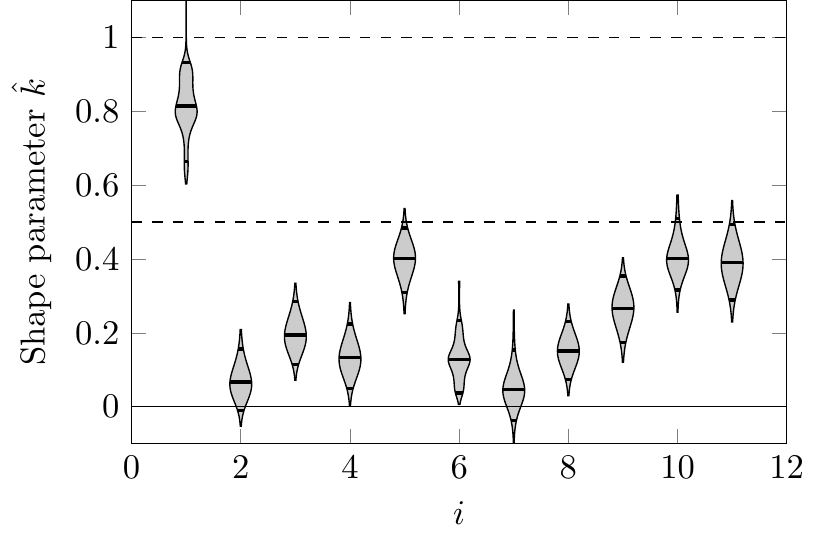}\hspace{.5cm}
  \includegraphics[width=.46\textwidth]{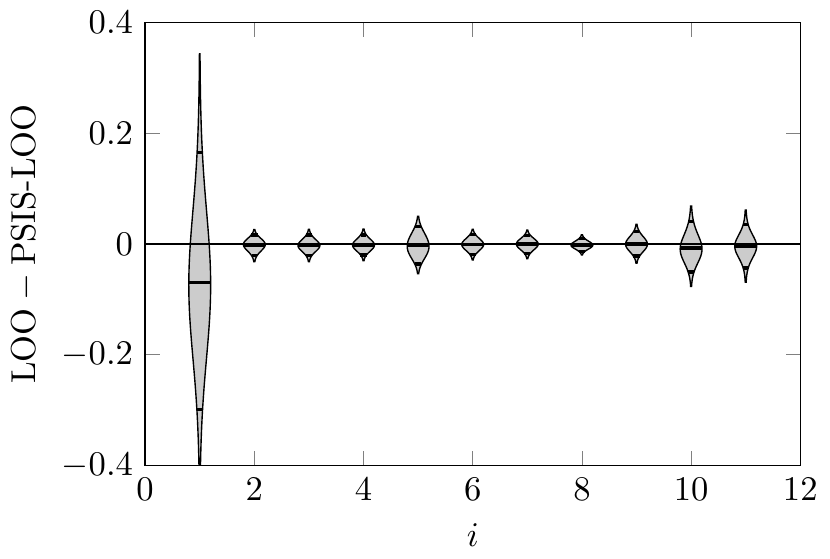}
}
\caption{\em Puromycin example:  Distributions of (a) tail shape estimates and
(b) PSIS-LOO estimation errors compared to LOO, from 100 independent Stan runs.
In an applied example we would only perform these calculations once, but here we
replicate 100 times to give a sense of the Monte Carlo error of our procedure.}
\label{fig:mm}
\end{figure}

Figure \ref{fig:mm} shows the distribution of the estimated tail shapes $k$ and
estimation errors compared to LOO in 100 independent Stan runs. The estimates of
the tail shape $k$ for $i=1$ suggest that the variance of the raw importance
ratios is infinite. However, the generalized central limit theorem for stable
distributions still holds and we can get an accurate estimate of the
corresponding term in LOO. We could obtain more draws to reduce the Monte Carlo
error, or again consider a more robust model.

\subsection{Example: Logistic regression for leukemia survival}\label{leukemia}

Our next example uses a logistic regression model to predict survival of
leukemia patients past 50 weeks from diagnosis. These data were also analyzed by
Epifani et al.\ (2008). Explanatory variables are white blood cell count at
diagnosis and whether ``Auer rods and/or significant granulature of the leukemic
cells in the bone marrow at diagnosis'' were present.

\begin{figure}
\centerline{
   \includegraphics[width=.45\textwidth]{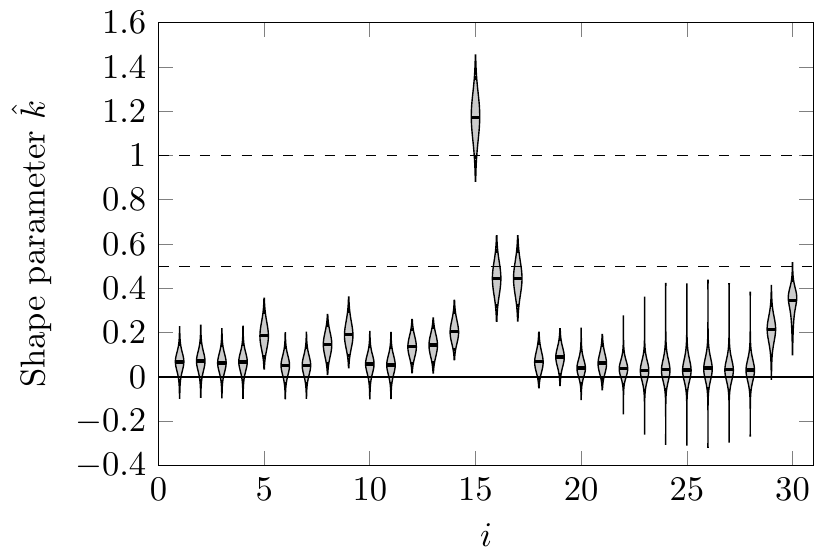}\hspace{.5cm}
  \includegraphics[width=.44\textwidth]{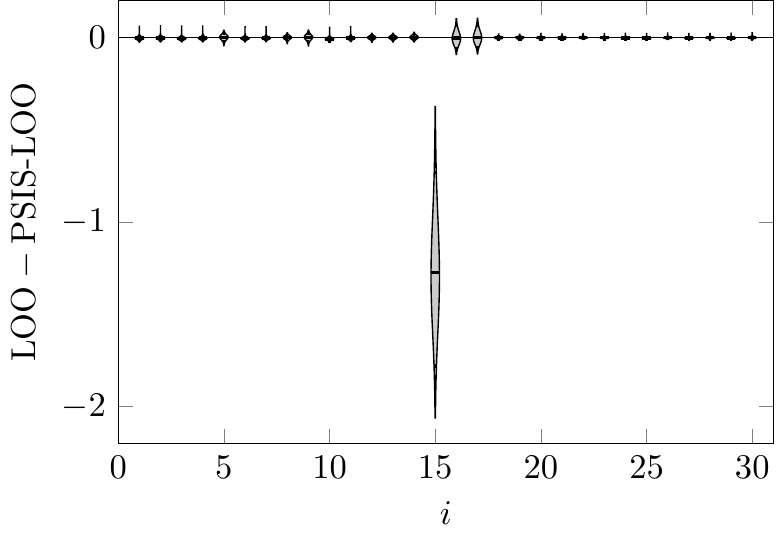}
}
\caption{\em Leukemia example:  Distributions of (a) tail shape estimates and
(b) PSIS-LOO estimation errors compared to LOO, from 100 independent Stan runs.
The pointwise calculation of the terms in PSIS-LOO reveals that much of the
uncertainty comes from a single data point, and it could make sense to simply
re-fit the model to the subset and compute LOO directly for that point.}
\label{fig:leukemiax1}
\end{figure}

Epifani et al.\ (2008) show that the raw importance ratios for data point $i=15$
have infinite variance. Figure \ref{fig:leukemiax1} shows the distribution of
the estimated tail shapes $k$ and estimation errors compared to LOO in 100
independent Stan runs. The estimates of the tail shape $k$ for $i=15$ suggest
that the mean and variance of the raw importance ratios do not exist. Thus the
generalized central limit theorem does not hold.

\begin{figure}
\centerline{
   \includegraphics[width=.45\textwidth]{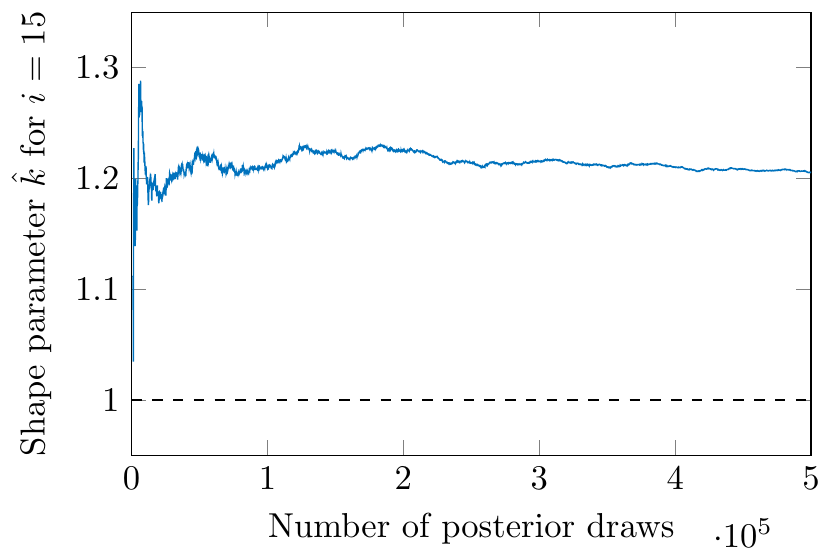}\hspace{.5cm}
  \includegraphics[width=.45\textwidth]{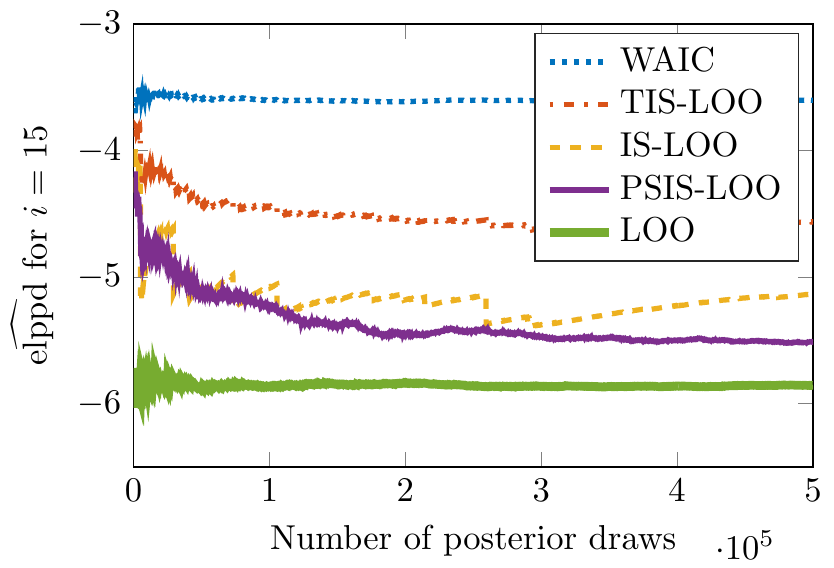}
}
\caption{\em Leukemia example:  Distributions of (a) tail shape estimate and (b)
LOO approximations for $i=15$.  If we continue sampling, the tail shape estimate
stays above 1 and $\widehat{\rm elpd}_i$ will not converge.}
\label{fig:leukemiax_15}
\end{figure}

Figure \ref{fig:leukemiax_15} shows that if we continue sampling, the tail shape
estimate stays above 1 and $\widehat{\rm elpd}_i$ will not converge.

\begin{figure}
\centerline{
   \includegraphics[width=.45\textwidth]{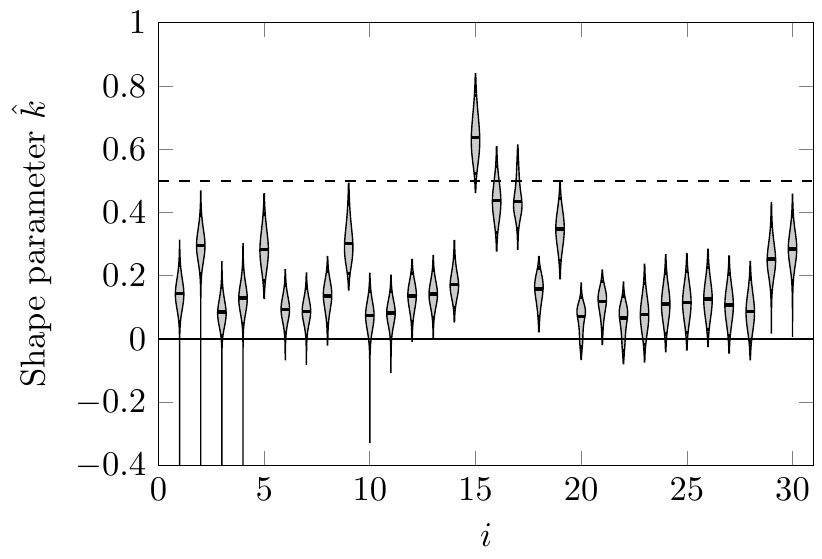}\hspace{.5cm}
  \includegraphics[width=.45\textwidth]{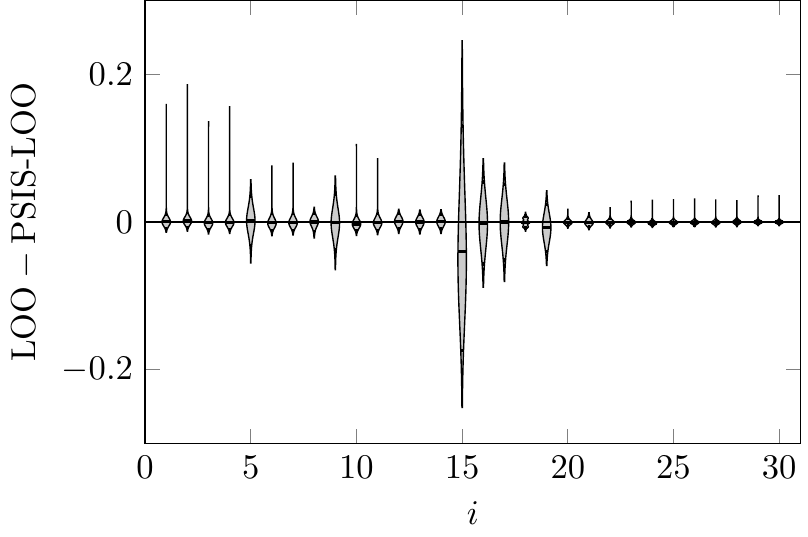}
}
\caption{\em Leukemia example with log-transformed predictor: (a)
Distributions of tail shape estimates for each data point and (b) PSIS-LOO estimation errors
compared to LOO, from 100 independent Stan runs.  Computations are more stable
compared to the model fit on the original scale and displayed in Figure
\ref{fig:leukemiax1}.}
\label{fig:leukemiax2}
\end{figure}

Large estimates for the tail shape parameter indicate that the full posterior is
not a good importance sampling approximation for the desired leave-one-out
posterior, and thus the observation is surprising. The original model used the
white blood cell count directly as a predictor, and it would be natural to use
its logarithm instead. Figure \ref{fig:leukemiax2} shows the distribution of the
estimated tail shapes $k$ and estimation errors compared to LOO in 100
independent Stan runs for this modified model. Both the tail shape values and
errors are now smaller.

\subsection{Example: Multilevel regression for radon contamination}\label{radon}

Gelman and Hill (2007) describe a study conducted by the United States
Environmental Protection Agency designed to measure levels of the carcinogen
radon in houses throughout the United States. In high concentrations radon is
known to cause lung cancer and is estimated to be responsible for several
thousands of deaths every year in the United States. Here we focus on the sample
of 919 houses in the state of Minnesota, which are distributed (unevenly)
throughout 85 counties.

We fit the following multilevel linear model to the radon data
\begin{align*}
y_i & \sim {\rm N}\left(\alpha_{j[i]} + \beta_{j[i]} x_i, \sigma^2\right), \quad i = 1, \ldots, 919 \\[5pt]
\begin{pmatrix} \alpha_j \\ \beta_j \end{pmatrix} & \sim {\rm N} \left(
  \begin{pmatrix}
	\gamma_0^\alpha + \gamma_1^\alpha u_j \\
	\gamma_0^\beta + \gamma_1^\beta u_j
  \end{pmatrix},
  \begin{pmatrix}
	\sigma^2_\alpha & \rho \sigma_\alpha \sigma_\beta  \\
 	\rho \sigma_\alpha \sigma_\beta & \sigma^2_\beta
  \end{pmatrix}
\right), \quad j = 1, \ldots, 85,
\end{align*}
where $y_i$ is the logarithm of the radon measurement in the $i$th house, $x_i =
0$ for a measurement made in the basement and $x_i  = 1$ if on the first floor
(it is known that radon enters more easily when a house is built into the
ground), and the county-level predictor $u_j$ is the logarithm of the soil
uranium level in the county. The residual standard deviation $\sigma$ and all
hyperparameters are given weakly informative priors. Code for fitting this 
model is provided in Appendix~\ref{sec:rstanarm}.

The sample size in this example $(n=919)$ is not huge but is large enough that
it is important to have a computational method for LOO that is fast for each
data point. Although the MCMC for the full posterior inference (using four
parallel chains) finished in only 93 seconds, the computations for exact brute
force LOO require fitting the model 919 times and took more than 20 hours to
complete (Macbook Pro, 2.6 GHz Intel Core i7). With the same hardware the
PSIS-LOO computations took less than 5 seconds.

\begin{figure}
\centerline{
   \includegraphics[width=.45\textwidth]{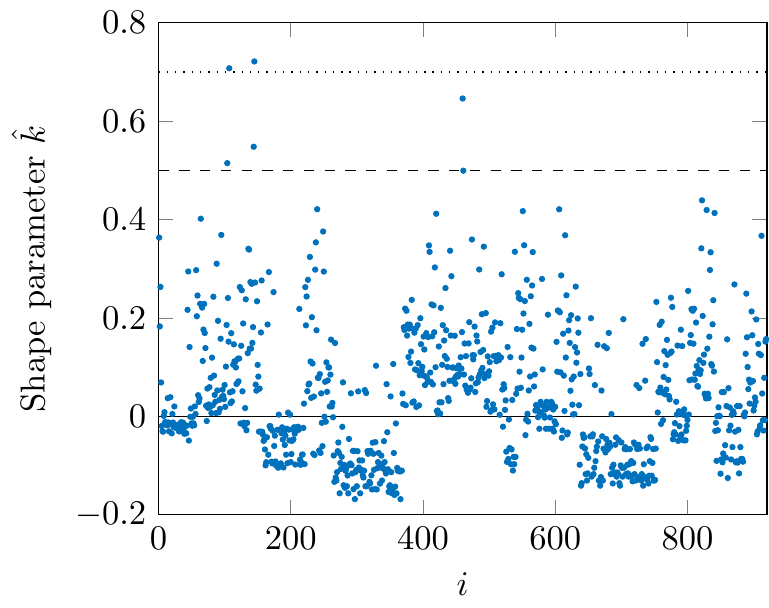}\hspace{.5cm}
  \includegraphics[width=.45\textwidth]{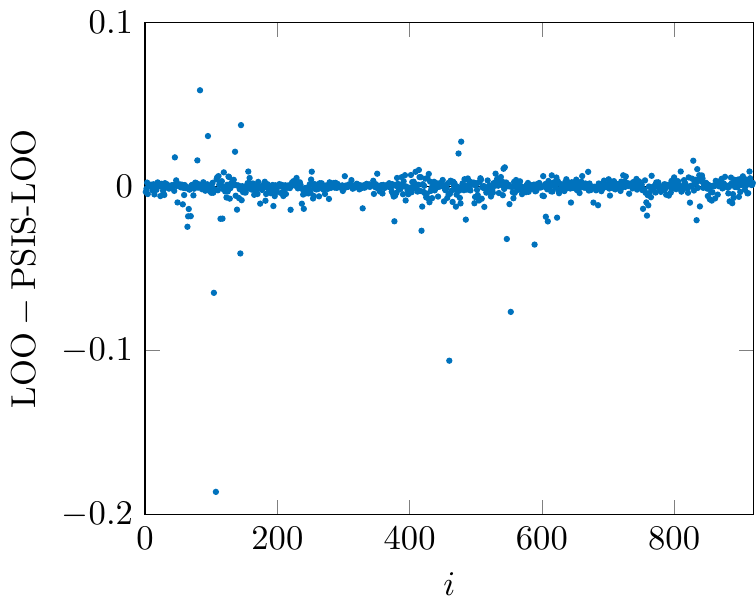}
}
\caption{\em Radon example:  (a) Tail shape estimates for each point's contribution
to LOO, and (b) error in PSIS-LOO accuracy for each data point, all based on a single fit of the model in Stan.}
\label{fig:radon1}
\end{figure}

Figure \ref{fig:radon1} shows the results for the radon example and indeed the estimated 
shape parameters $k$ are small and all of the tested methods are accurate. For two 
observations the estimate of $k$ is slightly higher than the preferred threshold of $0.7$, 
but we can easily compute the elpd contributions for these points directly and then combine 
with the PSIS-LOO estimates for the remaining observations.\footnote{%
As expected, the two slightly high estimates for $k$ correspond to particularly influential 
observations, in this case houses with extremely low radon measurements.}
This is the procedure we refer to as PSIS-LOO+ in Section~\ref{examples-summary} below.

\begin{table}
  \centering
\begin{small}
  \begin{tabular}{l r r r r r r r}
    method & 8 schools & Stacks-$N$ & Stacks-$t$ & Puromycin & Leukemia & Leukemia-log & Radon\\ \hline
    PSIS-LOO  & 0.21 & 0.21 & 0.12 & 0.20 & 1.33 & 0.18 & 0.34 \\
    IS-LOO    & 0.28 & 0.37 & 0.12 & 0.28 & 1.43 & 0.21 & 0.39 \\
    TIS-LOO   & 0.19 & 0.37 & 0.12 & 0.27 & 1.80 & 0.18 & 0.36 \\
    WAIC      & 0.40 & 0.68 & 0.12 & 0.46 & 2.30 & 0.29 & 1.30 \\
    PSIS-LOO+ & 0.21 & 0.11 & 0.12 & 0.10 & 0.11 & 0.18 & 0.34 \\
    $10$-fold-CV &$-$& 1.34 & 1.01 & $-$  & 1.62 & 1.40 & 2.87 \\
    $10\times10$-fold-CV
              & $-$  & 0.46 & 0.38 & $-$  & 0.43 & 0.36 & $-$
  \end{tabular}
\end{small}
\caption{\em Root mean square error for different computations of LOO as
determined from a simulation study, in each case based on running Stan to obtain
4000 posterior draws and repeating 100 times.  Methods compared are Pareto smoothed importance sampling
(PSIS), PSIS with direct sampling if $\hat{k}_i>0.7$ (PSIS-LOO+), raw importance
sampling (IS), truncated importance sampling (TIS), WAIC, $10$-fold-CV, and $10$
times repeated $10$-fold-CV for the different examples considered in Sections
\ref{schools}--\ref{radon}: the hierarchical model for the 8 schools, the stack
loss regression (with normal and $t$ models), nonlinear regression for
Puromycin, logistic regression for leukemia (in original and log scale), and hierarchical linear regression for radon. See text for explanations. PSIS-LOO and
PSIS-LOO+ give the smallest error in all examples except the 8 schools, where it
gives the second smallest error. In each case, we compared the estimates to the
correct value of LOO by the brute-force procedure of fitting the model
separately to each of the $n$ possible training sets for each example.}
  \label{tab:rmse}
\end{table}

\begin{table}\small
  \centering
\begin{small}
  \begin{tabular}{l r r r r r r r}
    method  & 8 schools & Stacks-$N$ & Stacks-$t$ & Puromycin & Leukemia & Leukemia-log & Radon \\ \hline
    PSIS-LOO  & 0.19 & 0.12 & 0.07 & 0.10 & 1.02 & 0.09 & 0.18 \\
    IS-LOO  & 0.13 & 0.21 & 0.07 & 0.25 & 1.21 & 0.11 & 0.24 \\
    TIS-LOO & 0.15 & 0.27 & 0.07 & 0.17 & 1.60 & 0.09 & 0.24 \\
    WAIC    & 0.40 & 0.67 & 0.09 & 0.44 & 2.27 & 0.25 & 1.30
  \end{tabular}
\end{small}
  \caption{\em Partial replication of Table \ref{tab:rmse} using 16,000 posterior draws
  in each case. Monte Carlo errors are slightly lower.  The errors for WAIC do
  not simply scale with $1/\sqrt{S}$ because most of its errors come from bias
  not variance.}
  \label{tab:rmse16000}
\end{table}

\subsection{Summary of examples}
\label{examples-summary}

Table \ref{tab:rmse} compares the performance of Pareto smoothed importance
sampling, raw importance sampling, truncated importance sampling, and
WAIC for estimating expected out-of-sample prediction accuracy for each of the
examples in Sections \ref{schools}--\ref{radon}. Models were fit in Stan to
obtain 4000 simulation draws. In each case, the distributions come from 100
independent simulations of the entire fitting process, and the root mean squared
error is evaluated by comparing to exact LOO, which was computed by separately
fitting the model to each leave-one-out dataset for each example.
The last three lines of Table \ref{tab:rmse} show additionally the performance
of PSIS-LOO combined with direct sampling for the problematic $i$ with
$\hat{k}>0.7$ (PSIS-LOO+), $10$-fold-CV, and $10$ times repeated $10$-fold-CV.\footnote{$10$-fold-CV results were not computed for data sets with $n\leq 11$, and $10$ times
repeated $10$-fold-CV was not feasible for the radon example due to the computation
time required.}
For the Stacks-$N$, Puromycin, and Leukemia examples, there was one $i$ with
$\hat{k}>0.7$, and thus the improvement has the same computational cost as the
full posterior inference. $10$-fold-CV has higher RMSE than LOO approximations
except in the Leukemia case. The higher RMSE of $10$-fold-CV is due to
additional variance from the data division. The repeated $10$-fold-CV has
smaller RMSE than basic $10$-fold-CV, but now the cost of computation is already 
100 times the original full posterior inference. These results show that $K$-fold-CV 
is needed only if LOO approximations fail badly (see also the results in 
Vehtari \& Lampinen, 2002).

As measured by root mean squared error, PSIS consistently performs well. In
general, when IS-LOO has problems it is because of the high variance of the raw
importance weights, while TIS-LOO and WAIC have problems because of bias. Table
\ref{tab:rmse16000} shows a replication using 16,000 Stan draws for each
example. The results are similar results and PSIS-LOO is able to improve the
most given additional draws.

\section{Standard errors and model comparison}\label{diagnostics}
We next consider some approaches for assessing the uncertainty of
cross-validation and WAIC estimates of prediction error.  We present these
methods in a separate section rather than in our main development because, as
discussed below, the diagnostics can be difficult to interpret when the sample
size is small.

\subsection{Standard errors}
The computed estimates $\widehat{\mbox{elpd}}_{\rm loo}$ and
$\widehat{\mbox{elpd}}_{\rm waic}$ are each defined as the sum of $n$
independent components so it is trivial to compute their standard errors by
computing the standard deviation of the $n$ components and multiplying by
$\sqrt{n}$. For example, define
\begin{equation}
 \widehat{\mbox{elpd}}_{{\rm loo},i} = \log p(y_i|y_{-i}),
\end{equation}
so that $\widehat{\mbox{elpd}}_{\rm loo}$ in (\ref{xformula2}) is the sum of
these $n$ independent terms. Then
\begin{equation}\label{se.loo}
\mbox{se}\,(\widehat{\mbox{elpd}}_{\rm loo}) =
\sqrt{n\,{\mathlarger V}_{i=1}^n \widehat{\rm elpd}_{{\rm loo},i}},
\end{equation}
and similarly for WAIC and $K$-fold cross-validation. The effective numbers of
parameters, $\widehat{p}_{\rm loo}$ and $ \widehat{p}_{\rm waic}$, are also sums
of independent terms so we can compute their standard errors in the same way.

These standard errors come from considering the $n$ data points 
as a sample from a larger population or, equivalently, as independent
realizations of an error model. One can also compute Monte Carlo standard errors
arising from the finite number of simulation draws using the formula from Gelman
et al.\ (2013) which uses both between and within-chain information and is
implemented in Stan. In practice we expect Monte Carlo standard errors to not be
so interesting because we would hope to have enough simulations that the
computations are stable, but it could make sense to look at them just to check
that they are low enough to be negligible compared to sampling error (which
scales like $1/n$ rather than $1/S$).

The standard error (\ref{se.loo}) and the corresponding formula for
$\mbox{se}\,(\widehat{\mbox{elpd}}_{\rm waic})$ have two difficulties when
the sample size is low.  First, the $n$ terms are not strictly independent because
they are all computed from the same set of posterior simulations $\theta^s$.
This is a generic issue when evaluating the standard error of any
cross-validated estimate. Second, the terms in any of these expressions can come
from highly skewed distributions, so the second moment might not give a good
summary of uncertainty.  Both of these problems should subside as $n$ becomes
large.  For small $n$, one could instead compute nonparametric error estimates
using a Bayesian bootstrap on the computed log-likelihood values corresponding
to the $n$ data points (Vehtari and Lampinen, 2002).

\subsection{Model comparison}
When comparing two fitted models, we can estimate the difference in their
expected predictive accuracy by the difference in $\widehat{\mbox{elpd}}_{\rm
loo}$ or $\widehat{\mbox{elpd}}_{\rm waic}$ (multiplied by $-2$, if desired, to
be on the deviance scale). To compute the standard error of this difference we
can use a paired estimate to take advantage of the fact that the same set of $n$
data points is being used to fit both models.

For example, suppose we are comparing models A and B, with corresponding fit
measures
$\widehat{\rm elpd}_{\rm loo}^A\!=\!\sum_{i=1}^n \widehat{\rm elpd}_{{\rm loo},i}^A$
and
$\widehat{\rm elpd}_{\rm loo}^B\!=\!\sum_{i=1}^n \widehat{\rm elpd}_{{\rm loo},i}^B$.
The standard error of their difference is simply,
\begin{equation}\label{se.diff}
\mbox{se}\,(\widehat{\mbox{elpd}}_{\rm loo}^A- \widehat{\mbox{elpd}}_{\rm loo}^B)
=\sqrt{n\,{\mathlarger V}_{i=1}^n (\widehat{\rm elpd}_{{\rm loo},i}^A- \widehat{\rm elpd}_{{\rm loo},i}^B)},
\end{equation}
and similarly for WAIC and $K$-fold cross-validation. Alternatively the non-parametric 
Bayesian bootstrap approach can be used (Vehtari and Lampinen, 2002).

As before, these calculations should be most useful when $n$
is large, because then non-normality of the distribution is not such an issue
when estimating the uncertainty of these sums.

In any case, we suspect that these standard error formulas, for all their flaws,
should give a better sense of uncertainty than what is obtained using the
current standard approach for comparing differences of deviances to a $\chi^2$
distribution, a practice that is derived for Gaussian linear models or
asymptotically and, in any case, only applies to nested models.

Further research needs to be done to evaluate the performance in model
comparison of (\ref{se.diff}) and the corresponding standard error formula for
LOO. Cross-validation and WAIC should not be used to select a single model among
a large number of models due to a selection induced bias as demonstrated, for
example, by Piironen and Vehtari (2016).

We demonstrate the practical use of LOO in model comparison using the radon
example from Section~\ref{radon}. Model A is the multilevel linear model
discussed in Section \ref{radon} and Model B is the same model but without the
county-level uranium predictor. That is, at the county-level Model B has
$$
\begin{pmatrix} \alpha_j \\ \beta_j \end{pmatrix} \sim
{\rm N} \left(
  \begin{pmatrix}
	\mu_\alpha \\
	\mu_\beta
  \end{pmatrix},
  \begin{pmatrix}
	\sigma^2_\alpha & \rho \sigma_\alpha \sigma_\beta  \\
 	\rho \sigma_\alpha \sigma_\beta & \sigma^2_\beta
  \end{pmatrix}
\right), \quad j = 1, \ldots, 85.
$$
Comparing the models on PSIS-LOO reveals an estimated difference in elpd of
$10.2$ (with a standard error of $5.1$) in favor of Model A.

\subsection{Model comparison using pointwise prediction errors}\label{s:arsenic}

We can also compare models in their leave-one-out errors, point by point.  We illustrate with an analysis of a survey of residents from a small 
area in Bangladesh that was affected by arsenic in drinking water. Respondents with elevated 
arsenic levels in their wells were asked if they were interested in getting water from a neighbor's well, 
and a series of models were fit to predict this binary response given various information 
about the households (Gelman and Hill, 2007).

Here we start with a logistic regression for the well-switching response 
given two predictors: the arsenic level of the water in the resident's home, and the distance of the 
house from the nearest safe well.  We compare this to an alternative logistic regression with the arsenic predictor on the logarithmic scale.
The two models have the same number of parameters but give different predictions.

\begin{figure}
\centerline{
   \includegraphics[width=.42\textwidth]{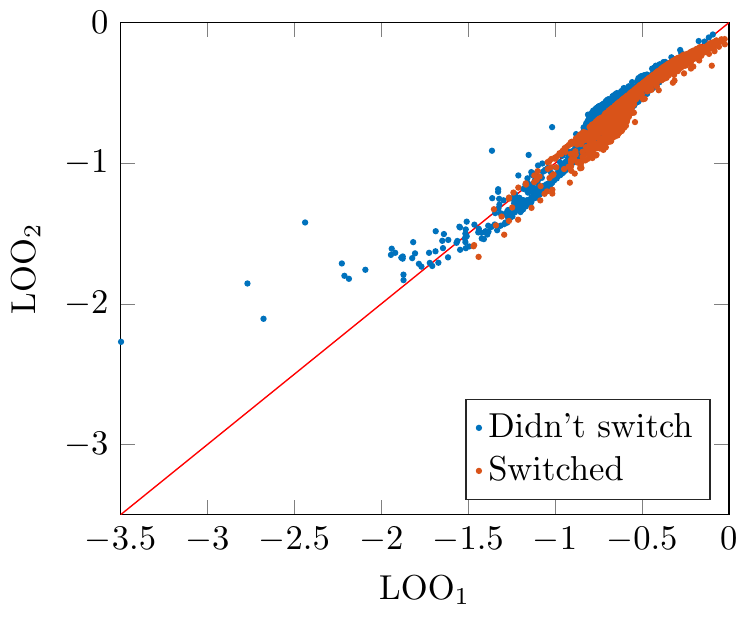}\hspace{1cm}
  \includegraphics[width=.438\textwidth]{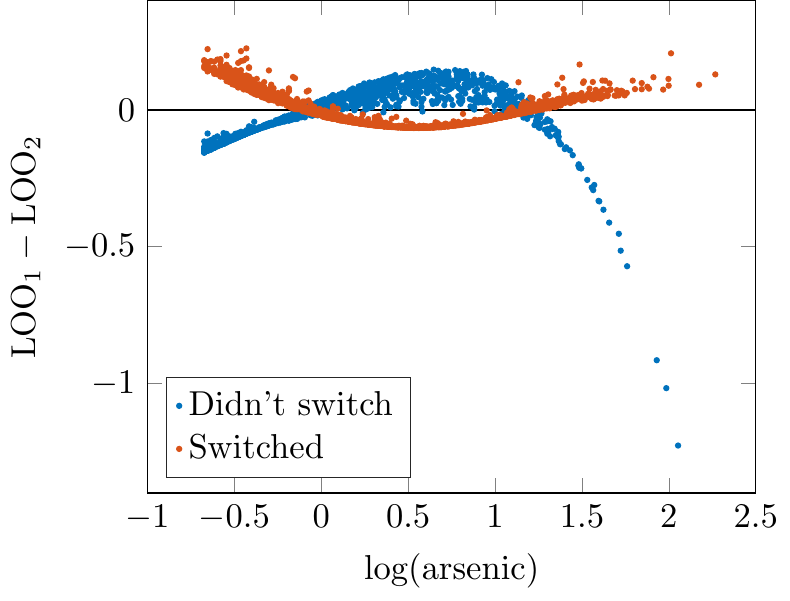}
}
\caption{\em Arsenic example, comparing two models in terms of their pointwise contributions to LOO:  
(a) comparing contributions of LOO directly; (b) plotting the difference in LOO as a function of a key predictor 
(the existing arsenic level).  To aid insight, we have colored the data according to the (binary) output, with red
corresponding to $y=1$ and blue representing $y=0$.  For any given data point, one model will fit better than 
another, but for this example the graphs reveal that the difference in LOO between the models arises from 
the linear model's poor predictions for 10--15 non-switchers with high arsenic levels.}
\label{fig:arsenic2}
\end{figure}

Figure \ref{fig:arsenic2} shows the pointwise results for the arsenic example. The scattered blue dots on 
the left side of Figure \ref{fig:arsenic2}a and on the lower right of Figure \ref{fig:arsenic2}b correspond to 
data points which Model A fits particularly poorly---that is, large negative contributions to the expected log predictive density.  We can also sum these $n$ terms to yield an estimated difference in 
${\rm elpd}_{\rm loo}$ of 16.4 with a standard error of 4.4.  This standard error derives from the finite sample 
size and is scaled by the variation in the differences displayed in Figure \ref{fig:arsenic2}; it is {\em not} a 
Monte Carlo error and does not decline to 0 as the number of Stan simulation draws increases.

\section{Discussion}

This paper has focused on the practicalities of implementing LOO, WAIC, and
$K$-fold cross-validation within a Bayesian simulation environment, in
particular the coding of the log-likelihood in the model, the computations of
the information measures, and the stabilization of weights to enable an
approximation of LOO without requiring refitting the model.

Some difficulties persist, however.  As discussed above, any predictive accuracy
measure involves two definitions:  (1) the choice of what part of the model to
label as ``the likelihood,'' which is directly connected to which potential
replications are being considered for out-of-sample prediction; and (2) the
factorization of the likelihood into ``data points,'' which is reflected in the
later calculations of expected log predictive density.

Some choices of replication can seem natural for a particular
dataset but less so in other comparable settings.  For example, the 8 schools
data are available only at the school level and so it seems natural to treat the
school-level estimates as data.  But if the original data had been available, we
would surely have defined the likelihood based on the individual students' test
scores. It is an awkward feature of predictive error measures that they might be
determined based on computational convenience or data availability rather than
fundamental features of the problem. To put it another way, we are assessing the fit of the model to the particular data at hand.

Finally, these methods all have limitations. The concern with WAIC is that
formula (\ref{cp0}) is an asymptotic expression for the bias of lpd for
estimating out-of-sample prediction error and is only an approximation for
finite samples.  Cross-validation (whether calculated directly by re-fitting the
model to several different data subsets, or approximated using importance
sampling as we did for LOO) has a different problem in that it relies on
inference from a smaller subset of the data being close to inference from the
full dataset, an assumption that is typically but not always true.

For example, as we demonstrated in Section \ref{schools}, in a hierarchical
model with only one data point per group, PSIS-LOO and WAIC can dramatically
understate prediction accuracy.  Another setting where LOO (and cross-validation
more generally) can fail is in models with weak priors and sparse data.  For
example, consider logistic regression with flat priors on the coefficients and
data that happen to be so close to separation that the removal of a single data
point can induce separation and thus infinite parameter estimates.  In this case
the LOO estimate of average prediction accuracy will be zero (that is,
$\widehat{\mbox{elpd}}_{\rm is-loo}$ will be $-\infty$) if it is calculated to
full precision, even though predictions of future data from the actual fitted
model will have bounded loss.  Such problems should not arise asymptotically
with a fixed model and increasing sample size but can occur with actual finite
data, especially in settings where models are increasing in complexity and are
insufficiently constrained.

That said, quick estimates of out-of-sample prediction error can be valuable for
summarizing and comparing models, as can be seen from the popularity of AIC and
DIC.  For Bayesian models, we prefer PSIS-LOO and K-fold cross-validation to
those approximations which are based on point estimation.

\section*{References}

\noindent

\bibitem Akaike, H. (1973).  Information theory and an extension of the maximum likelihood principle.  In {\em Proceedings of the Second International Symposium on Information Theory}, ed.\ B. N. Petrov and F. Csaki, 267--281.  Budapest:  Akademiai Kiado.  

\bibitem Ando, T., and Tsay, R. (2010).  Predictive likelihood for Bayesian model selection and averaging. {\em International Journal of Forecasting} {\bf 26}, 744--763.

\bibitem Arlot, S., and Celisse, A. (2010). A survey of
  cross-validation procedures for model selection. {\em Statistics
    Surveys} {\bf 4}, 40--79.

\bibitem Bernardo, J., and Smith A. F. M (1994). {\em Bayesian theory}. Wiley.

\bibitem Burman, P. (1989). A comparative study of ordinary cross-validation, $v$-fold cross-validation and the repeated learning-testing methods. {\em Biometrika} {\bf 76},
503--514.

\bibitem Epifani, I., MacEachern, S. N., and Peruggia, M. (2008). Case-deletion importance sampling esti\-ma\-tors: Central limit theorems and related results. {\em Electronic Journal of Statistics} {\bf 2}, 774--806.

\bibitem Gabry, J., and Goodrich, B. (2016). rstanarm: Bayesian applied regression modeling via Stan.
R package version 2.10.0. \url{http://mc-stan.org/interfaces/rstanarm}

\bibitem Geisser, S., and Eddy, W. (1979). A predictive approach to model selection. {\em Journal of the American Statistical Association} {\bf 74}, 153--160.

\bibitem Gelfand, A. E. (1996).  Model determination using sampling-based methods.  In {\em Markov Chain Monte Carlo in Practice}, ed.\ W. R. Gilks, S. Richardson, D. J. Spiegelhalter, 145--162. London: Chapman and Hall.

\bibitem Gelfand, A. E., Dey, D. K., and Chang, H. (1992).  Model determination using predictive distributions with implementation via sampling-based methods.  In {\em Bayesian Statistics 4}, ed.\ J. M. Bernardo, J. O. Berger, A. P. Dawid, and A. F. M. Smith, 147--167.  Oxford University Press.

\bibitem Gelman, A., Carlin, J. B., Stern, H. S., Dunson, D. B., Vehtari, A., and Rubin D. B. (2013). {\em Bayesian Data Analysis}, third edition.  London:  CRC Press.

\bibitem Gelman, A., and Hill, J. (2007).  {\em Data Analysis Using Regression and Multilevel/Hier\-ar\-chi\-cal Models}.  Cambridge University Press.

\bibitem Gelman, A., Hwang, J., and Vehtari, A. (2014).  Understanding predictive information criteria for Bayesian models.  {\em Statistics and Computing} {\bf 24}, 997--1016.

\bibitem Gneiting, T., and Raftery, A. E. (2007). Strictly proper scoring rules, prediction, and estimation. {\em Journal of the American Statistical Association} {\bf 102}, 359--378.

\bibitem Hoeting, J., Madigan, D., Raftery, A. E., and Volinsky, C. (1999).  Bayesian model averaging. {\em Statistical Science} {\bf 14}, 382--417.

\bibitem Hoffman, M. D., and Gelman, A. (2014). The no-U-turn sampler: Adaptively setting path lengths in Hamiltonian Monte Carlo. {\em Journal of Machine Learning Research} {\bf 15}, 1593−1623.

\bibitem Ionides, E. L. (2008). Truncated importance sampling. {\em Journal of Computational and Graphical Statistics} {\bf 17}, 295–-311.



\bibitem Koopman, S. J., Shephard, N., and Creal, D. (2009). Testing the assumptions behind importance sampling. {\em Journal of Econometrics} {\bf 149}, 2--11.

\bibitem Peruggia, M. (1997). On the variability of case-deletion importance sampling weights in the Bayesian linear model. {\em Journal of the American Statistical Association} {\bf 92}, 199--207.

\bibitem Piironen, J., and Vehtari, A. (2016). Comparison of Bayesian predictive methods for model selection. {\em Statistics and Computing}. In press. \url{http://link.springer.com/article/10.1007/s11222-016-9649-y}
  
\bibitem Plummer, M. (2008).  Penalized loss functions for Bayesian model comparison.  {\em Biostatistics} {\bf 9}, 523--539.

\bibitem R Core Team (2016). R: A language and environment for statistical computing.
  R Foundation for Statistical Computing, Vienna, Austria. \url{https://www.R-project.org/}

\bibitem Rubin, D. B. (1981).  Estimation in parallel randomized experiments.
{\em Journal of Educational Statistics} {\bf 6}, 377--401.

\bibitem Spiegelhalter, D. J., Best, N. G., Carlin, B. P., and van der Linde, A. (2002).  Bayesian measures of model complexity and fit. {\em Journal of the Royal Statistical Society B} {\bf 64}, 583--639.

\bibitem Spiegelhalter, D., Thomas, A., Best, N., Gilks, W., and Lunn, D. (1994, 2003).  BUGS:  Bayesian inference using Gibbs sampling.  MRC Biostatistics Unit, Cambridge, England.\\
{\tt http://www.mrc-bsu.cam.ac.uk/bugs/}

\bibitem Stan Development Team (2016a).  The Stan C++ Library, version 2.10.0.
\url{http://mc-stan.org/}

\bibitem Stan Development Team (2016b).  RStan: the R interface to Stan, version 2.10.1.\\
\url{http://mc-stan.org/interfaces/rstan.html}

\bibitem Stone, M. (1977). An asymptotic equivalence of choice of model cross-validation and Akaike's criterion. {\em Journal of the Royal Statistical Society B} {\bf 36}, 44--47.

\bibitem van der Linde, A. (2005).  DIC in variable selection.  {\em Statistica Neerlandica} {\bf 1}, 45--56.

\bibitem Vehtari, A., and Gelman, A. (2015).  Pareto smoothed importance sampling. \url{arXiv:1507.02646}.

\bibitem Vehtari, A., Gelman, A., and Gabry, J. (2016). loo: Efficient leave-one-out
  cross-validation and WAIC for Bayesian models. R package version 0.1.6.
  \url{https://github.com/stan-dev/loo}

\bibitem Vehtari, A., Mononen, T., Tolvanen, V., Sivula, T., and Winther, O. (2016). Bayesian leave-one-out cross-validation approximations for Gaussian latent variable models. {\em Journal of Machine Learning Research}, 17(103):1--38.
  
\bibitem Vehtari, A., and Lampinen, J. (2002). Bayesian model assessment and comparison using cross-validation predictive densities. {\em Neural Computation} {\bf 14}, 2439--2468.

\bibitem Vehtari, A., and Ojanen, J. (2012).  A survey of Bayesian predictive methods for model assessment, selection and comparison.  {\em Statistics Surveys} {\bf 6}, 142--228.

\bibitem Vehtari, A., and Riihim{\"a}ki, J. (2014).  Laplace approximation for logistic Gaussian process density estimation and regression. {\em Bayesian analysis}, {\bf 9}, 425-448.

\bibitem Watanabe, S. (2010).  Asymptotic equivalence of Bayes cross validation and widely applicable information criterion in singular learning theory. {\em Journal of Machine Learning Research} {\bf 11}, 3571--3594.

\bibitem Zhang, J., and Stephens, M. A. (2009). A new and efficient estimation method for the generalized Pareto distribution. {\em Technometrics} {\bf 51}, 316--325.

\appendix

\section{Implementation in Stan and R}
\label{sec:implementation}

\subsection{Stan code for computing and storing the pointwise log-likelihood}
\label{subsec:stancode}

We illustrate how to write Stan code that computes and stores the pointwise log-likelihood
using the arsenic example from Section \ref{s:arsenic}.  We save the program in the file \verb+logistic.stan+:

\begin{small}\begin{quotation}\noindent\vspace{-1.5\baselineskip}
\begin{Verbatim}
data {
  int N;
  int P;
  int<lower=0,upper=1> y[N];
  matrix[N,P] X;
}
parameters {
  vector[P] b;
}
model {
  b ~ normal(0,1);
  y ~ bernoulli_logit(X*b);
}
generated quantities {
  vector[N] log_lik;
  for (n in 1:N)
    log_lik[n] = bernoulli_logit_lpmf(y[n] | X[n]*b);
}
\end{Verbatim}
\end{quotation}\end{small}

We have defined the log-likelihood as a vector {\tt log\_lik} in the generated
quantities block so that the individual terms will be saved by Stan.\footnote{ 
The code in the generated quantities block is written using the new syntax 
introduced in Stan version 2.10.0.}
It would seem desirable to compute the terms of the log-likelihood directly without
requiring the repetition of code, perhaps by flagging the appropriate lines in
the model or by identifying the log likelihood as those lines in the model that
are defined relative to the data.  But there are so many ways of writing any
model in Stan---anything goes as long as it produces the correct log posterior
density, up to any arbitrary constant---that we cannot see any general way at
this time for computing LOO and WAIC without repeating the likelihood part of
the code.  The good news is that the additional computations are relatively
cheap: sitting as they do in the generated quantities block (rather than in the
transformed parameters and model blocks), the expressions for the terms of the
log posterior need only be computed once per saved iteration rather than once
per HMC leapfrog step, and no gradient calculations are required.

\subsection{The {\tt loo} R package for LOO and WAIC}

The {\tt loo} R package provides the functions {\tt loo()} and {\tt waic()} for
efficiently computing PSIS-LOO and WAIC for fitted Bayesian models using the
methods described in this paper.

These functions take as their argument an $S \times n$ log-likelihood matrix,
where $S$ is the size of the posterior sample (the number of retained draws) and
$n$ is the number of data points.\footnote{For models fit to  large datasets
it can be infeasible to store the entire log-likelihood matrix in memory. A
function for computing the log-likelihood from the data and posterior draws of
the relevant parameters may be specified instead of the log-likelihood matrix---the necessary data and draws are supplied as an additional argument---and
columns of the log-likelihood matrix are computed as needed. This requires less
memory than storing the entire log-likelihood matrix and allows {\tt loo} to be
used with much larger datasets.} The required  means and variances across
simulations are calculated and then used to compute the effective number of
parameters and LOO or WAIC.

The {\tt loo()} function returns $\widehat{\mbox{elpd}}_{\rm loo}$,
$\widehat{p}_{\rm loo}$, ${\rm looic} =-2\, \widehat{\mbox{elpd}}_{\rm loo}$ (to
provide the output on the conventional scale of ``deviance'' or AIC),\footnote{In statistics there is a tradition of looking at deviance,
while in computer science the log score is more popular, so we return both.} the
pointwise contributions of each of these measures, and standard
errors. The
{\tt waic()} function computes the analogous quantities for WAIC. Also returned
by the {\tt loo()} function is the estimated shape parameter $\hat{k}$ for the
generalized Pareto fit to the importance ratios for each leave-one-out
distribution. These computations could also be implemented directly in Stan C++,
perhaps following the rule that the calculations are performed if there is a
variable named {\tt log\_lik}. The {\tt loo} R package, however, is more general
and does not require that a model be fit using Stan, as long as an appropriate
log-likelihood matrix is supplied.

\paragraph{Using the {\tt loo} package.}
Below, we provide R code for preparing and running the logistic regression for
the arsenic example in Stan. After fitting the model we then use the {\tt loo}
package to compute LOO and WAIC.%
\footnote{The {\tt extract\_log\_lik}()
function used in the example is a convenience function for extracting the
log-likelihood matrix from a fitted Stan model provided that the user has
computed and stored the pointwise log-likelihood in their Stan program (see, for
example, the {\tt generated quantities} block in \ref{subsec:stancode}). The
argument {\tt parameter\_name} (defaulting to {\tt "log\_lik"}) can also be
supplied to indicate which parameter or generated quantity corresponds to the
log-likelihood.}

\begin{small}\begin{quotation}\noindent\vspace{-1.5\baselineskip}
\begin{Verbatim}
library("rstan")
library("loo")
# Read in and prepare the data
wells <- read.csv("wells.csv")
N <- nrow(wells)
X <- cbind(rep(1,N), wells$dist100, wells$arsenic)
y <- wells$y
P <- ncol(X)
# Fit the model with Stan
fit_1 <- stan("logistic.stan")
print(fit_1, "b")
# Compute LOO
log_lik_1 <- extract_log_lik(fit_1)
loo_1 <- loo(log_lik_1)
print(loo_1)
\end{Verbatim}
\end{quotation}\end{small}
The printed output shows $\widehat{\mbox{elpd}}_{\rm loo}$,
$\widehat{p}_{\rm loo}$, ${\rm looic}$, and their standard errors:

\begin{small}\begin{quotation}\noindent\vspace{-1.5\baselineskip}
\begin{Verbatim}
Computed from 4000 by 3020 log-likelihood matrix
         Estimate   SE
elpd_loo  -1968.3 15.6
p_loo         3.1  0.1
looic      3936.6 31.2
All Pareto k estimates OK (k < 0.5)
\end{Verbatim}
\end{quotation}\end{small}

By default, the estimates for the shape parameter $k$ of the generalized Pareto
distribution are also checked and a message is displayed informing the user if
any $\hat{k}$ are problematic (see the end of Section~\ref{sec:LOO}). In the
example above the message tells us that all of the estimates for $k$ are fine.
However, if any $\hat{k}$ were between $1/2$ and $1$ or greater than $1$ the
message would instead look something like this:

\begin{small}\begin{quotation}\noindent\vspace{-1.5\baselineskip}
\begin{Verbatim}
Warning messages:
1: 200 (7%) Pareto k estimates between 0.5 and 1
2: 85 (3%) Pareto k estimates greater than 1
\end{Verbatim}
\end{quotation}\end{small}
If there are any warnings then it can be useful to visualize the estimates to
check which data points correspond to the large $\hat{k}$ values. A plot 
of the $\hat{k}$ estimates can also be generated using {\tt plot(loo1)} and 
the list returned by the {\tt loo} function also contains the full vector of 
$\hat{k}$ values.

\paragraph{Model comparison.}
To compare this model to a second model on their values of LOO we can use the
{\tt compare} function:

\begin{small}\begin{quotation}\noindent\vspace{-1.5\baselineskip}
\begin{Verbatim}
# First fit a second model, using log(arsenic) instead of arsenic
X <- cbind(rep(1,N), wells$dist100, log(wells$arsenic))
P <- ncol(X)
fit_2 <- stan("logistic.stan")
print(fit_2, "b")
log_lik_2 <- extract_log_lik(fit_2)
loo_2 <- loo(log_lik_2)

# Compare the models
loo_diff <- compare(loo_1, loo_2)
print(loo_diff)
\end{Verbatim}
\end{quotation}\end{small}
This new object, {\tt loo\_diff}, contains the estimated difference of expected
leave-one-out prediction errors between the two models, along with the standard
error:

\begin{quotation}\noindent\vspace{-1.5\baselineskip}
\begin{Verbatim}
elpd_diff        SE
     16.1       4.4
\end{Verbatim}
\end{quotation}

\paragraph{Code for WAIC.} For WAIC the code is analogous and the objects
returned have the same structure (except there are no Pareto $k$ estimates). The
{\tt compare()} function can also be used to estimate the difference in WAIC
between two models:

\begin{small}\begin{quotation}\noindent\vspace{-1.5\baselineskip}
\begin{Verbatim}
waic_1 <- waic(log_lik_1)
waic_2 <- waic(log_lik_2)
waic_diff <- compare(waic_1, waic_2)
\end{Verbatim}
\end{quotation}\end{small}

\subsection{Using the {\tt loo} R package with {\tt rstanarm} models}
\label{sec:rstanarm}

Here we show how to fit the model for the radon example  from Section~\ref{radon} and carry 
out PSIS-LOO using the {\tt rstanarm} and {\tt loo} packages.

\begin{small}\begin{quotation}\noindent\vspace{-1.5\baselineskip}
\begin{Verbatim}
library("rstanarm")
# The subset of the radon data we need is included in rstanarm
data(radon)
# Fit the first model
modelA <- stan_lmer(
  log_radon ~ floor + log_uranium + floor:log_uranium + (1 + floor | county), 
  data = radon, 
  cores = 4, 
  iter = 2000, 
  chains = 4
)
# Fit the model without the county-level uranium predictor
modelB <- update(fitA, formula = log_radon ~ floor + (1 + floor | county))
\end{Verbatim}
\end{quotation}\end{small}
After fitting the models we can pass the fitted model objects {\tt modelA} and 
{\tt modelB} directly to {\tt rstanarm}'s {\tt loo} method and it will call the
necessary functions from the {\tt loo} package internally.

\begin{small}\begin{quotation}\noindent\vspace{-1.5\baselineskip}
\begin{Verbatim}
# Compare models
looA <- loo(modelA) 
looB <- loo(modelB)
compare(looA, looB)
\end{Verbatim}
\end{quotation}\end{small}
This returns:

\begin{small}\begin{quotation}\noindent\vspace{-1.5\baselineskip}
\begin{Verbatim}
elpd_diff        se 
    -10.2       5.2 
\end{Verbatim}
\end{quotation}\end{small}
If there are warnings about large values of the estimated Pareto shape parameter $\hat{k}$ for the importance ratios,
{\tt rstanarm} is also able to automatically carry out 
the procedure we call PSIS-LOO+ (see Section~\ref{examples-summary}). 
That is, {\tt rstanarm} can refit the model, leaving out these problematic observations 
one at a time and computing their elpd contributions directly. Then these values 
are combined with the results from PSIS-LOO for the other observations 
and returned to the user. We recommended this when there 
are only a few large $\hat{k}$ estimates. If there are many of them then 
we recommend $K$-fold cross-validation, which is also implemented 
in the latest release of {\tt rstanarm}.

\subsection{Stan code for $K$-fold cross-validation}

To implement $K$-fold cross-validation we repeatedly partition the data, with
each partition fitting the model to the training set and using it to predict the
holdout set.  The code for cross-validation does not look so generic because of
the need to repeatedly partition the data. However, in any particular example
the calculations are not difficult to implement, the main challenge being the
increase in computation time by roughly a factor of $K$. We recommend doing the
partitioning in R (or Python, or whichever data-processing environment is being
used) and then passing the training data and holdout data to Stan in two pieces.

Again we illustrate with the logistic regression for the arsenic example. We
start with the model from above, but we pass in both the training data
({\tt N\_t, y\_t, X\_t}) and the holdout set ({\tt N\_h, y\_h, X\_h}),
augmenting the data block accordingly.  We then alter the generated quantities
block to operate on the holdout data:

\begin{small}\begin{quotation}\noindent\vspace{-1.5\baselineskip}
\begin{Verbatim}
data {
  int P;                          // Number of regression predictors
  int N_t;                        // (Training) number of data points
  int<lower=0,upper=1> y_t[N_t];  // (Training) binary data
  matrix[N_t,P] X_t;              // (Training) predictors
  int N_h;                        // (Holdout)
  int y_h[N_h];                   // (Holdout)
  matrix[N_h,P] X_h;              // (Holdout)
  real a;
}
parameters {
  vector[P] b;
}
model {
  y_t ~ bernoulli_logit(X_t*b);
}
generated quantities {
  vector[N_t] log_lik_t;
  vector[N_h] log_lik_h;
  for (n in 1:N_t)
    log_lik_t[n] = bernoulli_logit_lpmf(y_t[n] | X_t[n]*b);
  for (n in 1:N_h)
    log_lik_h[n] = bernoulli_logit_lpmf(y_h[n] | X_h[n]*b);
}
\end{Verbatim}
\end{quotation}\end{small}
LOO could be also implemented in this way, setting $N_t$ to $N-1$ and $N_h$ to
1.  But, as discussed in the article, for large datasets it is more
practical to approximate LOO using importance sampling on the draws from the
posterior distribution fit to the entire dataset.

\end{document}